\documentclass[12pt]{article}

\usepackage[dvipdfmx]{graphicx}
\usepackage[square,comma,numbers,sort&compress]{natbib}
\usepackage[left=20mm,right=20mm,top=20mm,bottom=20mm]{geometry}
\usepackage{color}
\usepackage{amsmath}
\usepackage{amssymb}
\usepackage{braket}
\usepackage{slashed}
\usepackage{float}
\usepackage{cases}
\usepackage{multirow}
\usepackage{relsize}

\newcommand{\al}[1]{\begin{align}#1\end{align}}

\newcommand{\bp}{\begin{pmatrix}}
\newcommand{\ep}{\end{pmatrix}}
\newcommand{\bb}{\begin{bmatrix}}
\newcommand{\eb}{\end{bmatrix}}

\newcommand{\del}{\partial}

\newcommand{\GkI}{I}
\newcommand{\GkII}{I\hspace{-.1em}I}
\newcommand{\GkIII}{I\hspace{-.1em}I\hspace{-.1em}I}

\usepackage[normalem]{ulem}

\newcommand{\beq}{\begin{equation}}
\newcommand{\eeq}{\end{equation}}
\newcommand{\bea}{\begin{eqnarray}}
\newcommand{\eea}{\end{eqnarray}}

\newcommand{\fulltoday}{\number\day\space \ifcase\month\or
    January\or February\or March\or April\or May\or June\or
    July\or August\or September\or October\or November\or December\fi
    \space\number\year}

 \makeatletter
    
    \@addtoreset{equation}{section}
  \makeatother

\allowdisplaybreaks[3]

\usepackage{calc}
\newcounter{hours}\newcounter{minutes}
\makeatletter                   
\renewcommand*{\thehours}{\two@digits\c@hours}
\renewcommand*{\theminutes}{\two@digits\c@minutes}
\makeatother

\begin{document}

\begin{titlepage}
\renewcommand\thefootnote{\alph{footnote}}
		\mbox{}\hfill KIAS-P16066, KOBE-TH-16-05\\
\vspace{4mm}
\begin{center}
{\fontsize{22pt}{0pt}\selectfont \bf {
{{6d Dirac fermion on a rectangle;} \\[6pt] scrutinizing boundary conditions, \\[6pt] mode functions and spectrum}
}} \\
\vspace{8mm}
	{\fontsize{14pt}{0pt}\selectfont \bf
	Yukihiro Fujimoto,\,\footnote{
		E-mail: \tt {y-fujimoto@oita-ct.ac.jp}
		}}
	{}
	{\fontsize{14pt}{0pt}\selectfont \bf
	{Kouhei} Hasegawa,\,\footnote{
		E-mail: \tt kouhei@phys.sci.kobe-u.ac.jp
		}}
	\\[6pt]
	{\fontsize{14pt}{0pt}\selectfont \bf
	Kenji Nishiwaki,\,\footnote{
		E-mail: \tt nishiken@kias.re.kr
		}}
	{}
	{\fontsize{14pt}{0pt}\selectfont \bf
	Makoto Sakamoto,\,\,\footnote{
		E-mail: \tt dragon@kobe-u.ac.jp} }
	{}
	{\fontsize{14pt}{0pt}\selectfont \bf
	Kentaro Tatsumi\,\,\footnote{
		E-mail: \tt 134s110s@stu.kobe-u.ac.jp} } \\
\vspace{6mm}
	{\fontsize{13pt}{0pt}\selectfont
		${}^{\mathrm{a}}$\,\it {National Institute of Technology, Oita College}, Oita 870-0152, Japan
		\smallskip\\[3pt]
		${}^{\mathrm{b,\,d,\,e}}$\,\it Department of Physics, Kobe University, Kobe 657-8501, Japan
		\smallskip\\[3pt]
		${}^{\mathrm{c}}$\,\it School of Physics, Korea Institute for Advanced Study, Seoul 02455,
		Republic of Korea
		\smallskip\\
	}
\vspace{8mm}
{\normalsize \fulltoday}
\vspace{10mm}
\end{center}
\begin{abstract}
{\fontsize{12pt}{16pt}\selectfont{
We classify possible boundary conditions of a 6d Dirac fermion $\Psi$ on a rectangle under the requirement that the 4d Lorentz structure is maintained, and derive the profiles {and spectrum} of the zero modes and nonzero KK modes under the two specific boundary conditions, (i) 4d-chirality positive components being zero at the boundaries and (ii) internal chirality positive components being zero at the boundaries.
In the case of (i), twofold degenerated chiral zero modes appear which are localized towards specific directions of the rectangle pointed by an angle parameter $\theta$.
This leads to an implication for a new direction of pursuing the origin of three generations in the matter fields of the standard model, even though triple-degenerated zero modes are not realized in the six dimensions.
{When such 6d fermions couple with a 6d scalar with a vacuum expectation value, $\theta$ contributes to a mass matrix of zero-mode fermions consisting of Yukawa interactions.}
The emergence of the angle parameter $\theta$ originates from a rotational symmetry in the degenerated chiral zero modes on the rectangle extra dimensions since they do not feel the boundaries.
In the case of (ii), this rotational symmetry is promoted to the two-dimensional conformal symmetry though no chiral massless zero mode appears.
We also discuss the correspondence between our model on a rectangle and orbifold models in some details.
}}
\end{abstract}
\end{titlepage}
\renewcommand\thefootnote{\arabic{footnote}}
\setcounter{footnote}{0}

\newpage

\section{Introduction
\label{sec:introduction}}

Considering extra dimensions has been a fascinating direction for deriving better understandings on various aspects of the standard mode~(SM) especially during the past two decades, e.g. on the hierarchy problem between the electroweak scale and an ultraviolet scale~(see e.g. Refs.~\cite{ArkaniHamed:1998rs,Antoniadis:1998ig,Gogberashvili:1998vx,Randall:1999ee,Hatanaka:1998yp}).
In general, when hidden spacial directions exist in our universe, we should deliberate on the profiles of {fields} along the directions.
Here, {the} boundary conditions~(BCs) {of them} at the circumference of extra dimensions play a significant role in the determination of mode functions, especially in the zero modes which correspond to the lowest modes of the effective mass appearing after the Kaluza-Klein~(KK) decomposition among the extra spacial directions.
A powerful and widely used precept for determining {a class of} BCs is the variational principle.
An advantage of this method originates from the characteristic that we derive the equation of motion~(EOM) of higher dimensional fields and necessary conditions for BCs simultaneously.

When a five-dimensional~(5d) space {takes the direct product of the four dimensional~(4d) Minkowski spacetime times an interval}, general discussions are found e.g. in Refs.~\cite{Csaki:2003dt,Csaki:2003sh,Csaki:2005vy}, which include the possibility that fields contain boundary-localized interactions.
Another avenue for discussing BCs is to consider discrete parities around the fixed points of orbifolds~\cite{Kawamura:1999nj,Kawamura:2000ev,Hall:2001pg,Hebecker:2001jb}, which is not touched in this manuscript.
An interesting point on an interval is that the simple 5d Dirac mass term $M \overline{\Psi^{(\text{5d})}} \Psi^{(\text{5d})}$ is allowed and the sign and the magnitude of the mass parameter $M$ describe the localization of the (single) massless chiral zero mode of $\Psi^{(\text{5d})}$, which is realized when we adopt the chiral boundary conditions, $\Psi^{(\text{5d})}_{R} = 0$ or $\Psi^{(\text{5d})}_{L} = 0$, at both of {the two ends of an interval} (see e.g.~\cite{Fujimoto:2011kf}).\footnote{
On the $S^1/Z_2$ orbifold, the simple Dirac mass term is prohibited by the $Z_2$ symmetry.
On the other hand, $Z_2$-odd mass terms are written down consistently and chiral zero modes can be localized under the presence of them~\cite{Kaplan:2001ga}.
}
This mechanism is useful for creating the fermion mass hierarchies and mixing patterns observed in the SM in a natural sense~\cite{Fujimoto:2012wv,Fujimoto:2013ki,Fujimoto:2014fka,Cai:2015jla}.

Next, let us briefly look at the situation in six dimensions~(6d), where two spacial directions are {compactified} and more complicated structures of extra dimensions can be realized.
A fascinating aspect of two extra dimensions are the existence of nontrivial solitonic objects among the two directions, e.g. vortex~\cite{Libanov:2000uf,Frere:2000dc} (see also Refs.~\cite{Frere:2001ug,Frere:2003hn}) and constant quantized magnetic flux~\cite{Cremades:2004wa} (see also e.g. Refs.~\cite{Abe:2008fi,Abe:2008sx,Fujimoto:2013xha,Abe:2013bca,Abe:2014noa,Abe:2015yva,Matsumoto:2016okl,Fujimoto:2016zjs} with orbifolding), under which zero mode profiles become chiral, degenerated and quasi-localized.
Then, considering this direction can lead to a simultaneous explanation of the three features of the SM, {4d} chirality, three generations and mass hierarchies of the quarks and leptons.
Another interesting possibility is the spontaneous compactification on $S^2$~\cite{Horvath:1977st,RandjbarDaemi:1982hi} (see also e.g. Ref.~\cite{Dohi:2014fqa} as a review), where the radius is spontaneously stabilized by the cancellation between the curvature of $S^2$ and the contribution from the monopole configuration of the extra $U(1)_X$ gauge field in the Einstein-Maxwell equation describing the system.
Note that chiral zero modes are realizable when the $U(1)_X$ charges are suitably assigned.
Other types of discussions are found e.g. in Refs.~\cite{Neronov:2001qv,Parameswaran:2006db,Gogberashvili:2007gg}.\footnote{
See also Refs.~\cite{Kehagias:2004gy,Tran:2005vj,Nilse:2006jv,Cacciapaglia:2006tg,Cacciapaglia:2016xty,Andriot:2016rdd} for topics associated with boundary structures.
}

Here, a different viewpoint is important, which is difference in Lorentz structures, since it restrict possible Lorentz  representations of fermions in 5d and 6d.
As widely known, only Dirac fermions can exist in 5d, while both of Dirac and Weyl fermions are possible in 6d, where the chirality which discriminates 6d Weyl fermions is different from the 4d chirality (see section~\ref{sec:setup_and_bcs} for details).
A 6d Weyl fermion is decomposed into a pair of right and left-handed fermions with 4d chiralities, which is equivalent to a 4d Dirac fermion.
Thereby, as we touched beforehand, nontrivial backgrounds or singular points (e.g. fixed points of $T^2/Z_2$~\cite{Appelquist:2000nn}, $T^2/Z_4$~\cite{Dobrescu:2004zi,Burdman:2005sr}, $S^2/Z_2$~\cite{Maru:2009wu} orbifolds or identified points on real projective plane~($RP^2$)~\cite{Cacciapaglia:2009pa}, projective sphere~\cite{Dohi:2010vc}) are requested for realizations of chiral spectrum at the energy-lowest zero modes, where left or right zero modes are projected out.

In this paper, a generalized situation is scrutinized, when the 6d fermion is Dirac, which is equivalent to two 4d Dirac fermions.
For simplicity, we have discussions on a rectangle, which is a simple 6d generalization of an interval in 5d, without introducing nontrivial backgrounds (e.g. solitonic configurations among the extra dimensions) or boundary-localized terms at the leading order.\footnote{
Even in the absent case of boundary-localized interactions at the leading order, such kind of interactions are induced by loop corrections in general~\cite{Cheng:2002iz,DaRold:2003yi,Ponton:2005kx,Cacciapaglia:2011hx,Maru:2014cba}.
}
{An apparent difference between 5d and 6d is the number of the directions to impose the boundary conditions on. In 5d, only one direction exists for imposing the BCs, while in 6d, two directions are found.}

If the chiral boundary conditions, $\Psi_{R} (x,y) = 0$ or $\Psi_{L} (x,y) = 0$ at the boundaries, are allowed for a 6d Dirac fermion, occurrence of chiral zero modes itself is expected.
However, how many chiral modes are realized is rather unclear.
Besides, the forms of such zero modes would be nontrivial.
If multiple chiral zero modes can emerge, exploring this kind of directions would shed a new light on phenomenological model building in higher dimensional spacetime.\footnote{
Another possible motivation for discussing Dirac fermions is anomalies in higher dimensional spacetime (see e.g.~\cite{Dobrescu:2001ae,ArkaniHamed:2001is,Borghini:2001sa,Asaka:2002my,vonGersdorff:2003dt,Scrucca:2004jn,vonGersdorff:2006nt}).
Dirac fermions are less harmful than Weyl fermions since the bulk theory becomes vector-like (in a sense in higher dimensions).
}
Here, like the 5d case on an interval, the following 6d Dirac mass term $M \overline{\Psi} \Psi$ can be written down, where the mass parameter $M$ is expected to describe directions and magnitudes of localized profiles at the zero modes.

Also, contemplating possible BCs of a 6d Dirac fermion itself would be captivating since generalized BCs beyond the Neumann and Dirichlet ones bring us nontrivial phenomena.
For example, a complex scalar field on an interval under the Robin BCs at the two boundaries $y=0,L$ with two {length} parameters $L_\pm$, confining phases are observed at a part of the parameter space of $(L_+, L_-)$~\cite{Fujimoto:2011kf}.
Therefore, classifying the possible BCs of a 6d Dirac fermion can lead to a new properties of theories in extra dimension.\footnote{
One can refer to e.g. Refs.~\cite{Sakamoto:1999yk,Sakamoto:1999ym,Sakamoto:1999iv,Ohnishi:2000hs,Matsumoto:2001fp,Sakamoto:2001gn,Haba:2009uu} for seeing other aspects of boundary conditions.}

This paper is organized as follows.
In section~\ref{sec:setup_and_bcs}, after introducing the setup of a 6d Dirac fermion on a rectangle and looking at properties of 6d Lorentz structure, we derive the EOM and necessary conditions for BCs via the variational principle, and subsequently classify possible boundary conditions under the requirement that the 4d Lorentz structure is maintained.
In section~\ref{sec:KK_expansion}, we concentrate on analyzing the profiles of zero modes and non-zero KK modes in the two specific classes, (i) a 4d chiral mode is projected out at the boundaries ($\Psi_{L}(x,y) = 0$ at boundaries); (ii) a two-dimensional (2d) chirality among the rectangle is projected out at the boundaries ($\Psi_{+}(x,y) = 0$ at boundaries).
We note that analyzing all of possible configurations given in the classification in section~\ref{sec:KK_expansion} is beyond the scope of this paper, and the `opposite' cases, (i') $\Psi_{R}(x,y) = 0$ at boundaries; (ii') $\Psi_{-}(x,y) = 0$ at boundaries, are easily declared based on the knowledge of the two cases concretely discussed in this section.
{In section~\ref{sec:orbifold}, we construct corresponding cases by adopting the method of orbifold.}
In section~\ref{sec:misc}, we recapitulate the result obtained in {sections~\ref{sec:KK_expansion} and \ref{sec:orbifold},} and comment on it from various points of view.
Section~\ref{sec:summary} is devoted to summarizing the whole discussions made in this manuscript.
{In appendices~\ref{appendix:Z2Z2} and \ref{appendix:spinor_rotation}, we provide some details of calculations.}
{In appendix~\ref{sec:theta}, we show a concrete example that zero mode profiles of degenerate massless 4d chiral fermions can generate large mass splitting when (6d) fermions couple to a (6d) scalar with a vacuum expectation value~(VEV).}

\section{Setup and Possible Boundary Conditions
\label{sec:setup_and_bcs}}

\subsection{Setup of 6d Dirac Fermion
\label{sec:setup}}

At first in this section, we show our setup with mentioning details of adopted notations and useful formulas.
Let us start with a 6d Dirac fermion $\Psi$ whose action is given as
\al{
S = \int \! d^{4} x \int_{0}^{L_{1}} \!\!\!\! dy_{1} \int_{0}^{L_{2}} \!\!\!\! dy_{2} \,
	\overline{\Psi}(x,y) \left( i \Gamma^A \del_A - M \right) \Psi(x,y),
	\label{eq:6d_action}
}
where $x^\mu$ $(\mu = 0,1,2,3)$ denote the coordinate of the 4d Minkowski spacetime and $y_j$ $(j = 1,2)$ denote
the {coordinate} of the 2d extra space directions.
$\Psi(x,y)$ is an eight-component Dirac spinor on six dimensions.
The 2d extra space is taken to be a rectangle whose lengths of the two sides along the $y_1$ and $y_2$ axes are represented by $L_1$ and $L_2$, respectively.
In our convention, the six-dimensional Gamma matrices $\Gamma^A$ $(A = \mu,y_1,y_2)$ satisfy
\al{
\{ \Gamma^A, \Gamma^B \} &= -2 \eta^{AB} \, {\mathrm{I}_8} \quad (A, B = \mu,y_1,y_2), 
	\label{eq:Gamma_Clifford}\\
(\Gamma^A)^\dagger &= \begin{cases} +\Gamma^A & A=0, \\ -\Gamma^A & A \not= 0, \end{cases}
}
where the 6d metric is taken to be
\al{
\eta^{AB} = \eta_{AB} = \text{diag}(-1,1,1,1,1,1).
}
{$\mathrm{I}_n$ denotes the $n$-by-$n$ identity matrix.}
The Dirac conjugate $\overline{\Psi}$ is defined, as usual, by $\overline{\Psi} \equiv \Psi^\dagger \Gamma^0$.
The following relations easily derived from Eq.~(\ref{eq:Gamma_Clifford}) are useful
\al{
(\Gamma^A)^2 &= \begin{cases} + {\mathrm{I}_8} & A=0, \\ - {\mathrm{I}_8} & A \not= 0, \end{cases}\quad
(i \Gamma^{y_1})^2 = (i \Gamma^{y_2})^2 = {\mathrm{I}_8},
	\label{eq:Gamma_sq_relations}
}

As widely known, the degrees of freedom~(DOF) of a 6d Dirac spinor is equivalent to four 4d Weyl spinors.
Reflecting this fact, the two types of {\it chiralities}, $R/L$ and $+/-$, are defined as $+1$ or $-1$ eigenvalues of the following two matrices,
\al{
\Gamma^5 &\equiv i \Gamma^0 \Gamma^1 \Gamma^2 \Gamma^3,& {\Gamma^{y}} &\equiv i \Gamma^{y_1} \Gamma^{y_2}, \\
\Gamma^5 \Psi_{R\pm} &= + \Psi_{R\pm},&  \Gamma^5 \Psi_{L\pm} &= - \Psi_{L\pm}, \\
{\Gamma^{y}} \Psi_{R\pm} &= \pm \Psi_{R\pm},& {\Gamma^{y}} \Psi_{L\pm} &= \pm \Psi_{L\pm},
	\label{eq:eigenstate_Gammay3}
}
where the following relations are useful: $(\Gamma^5)^\dagger = \Gamma^5$, $({\Gamma^{y}})^\dagger = {\Gamma^{y}}$, $(\Gamma^5)^2 = ({\Gamma^{y}})^2 = {\mathrm{I}_8}$.
Here, the eigenvalues of $\Gamma^5$ and ${\Gamma^{y}}$ correspond to the ordinary 4d chirality and the {internal chirality} among the extra spacial directions~\cite{Green:87}, respectively.
The 6d chirality, which is defined as the eigenvalues of the following matrix, is expressed by use of the two eigenvalues as
\al{
\Gamma^7 \equiv - \Gamma^0 \Gamma^1 \Gamma^2 \Gamma^3 \Gamma^{y_1} \Gamma^{y_2} = \Gamma^5 \, {\Gamma^{y}}
	\label{eq:Gamma_7}
}
and the value is automatically determined.
The corresponding two types of projection matrices are defined in the usual manner,\footnote{
A set of projective operators into two states $s_1$ and $s_2$ {holds} the properties:
$(\mathcal{P}_{s_{1,2}})^2 = (\mathcal{P}_{s_{1,2}})$,
$\mathcal{P}_{s_1} \mathcal{P}_{s_2} = \mathcal{P}_{s_2} \mathcal{P}_{s_1} = 0$,
$\mathcal{P}_{s_1} + \mathcal{P}_{s_2} = 1$.
}
\al{
\mathcal{P}_{R/L} \equiv \frac{{\mathrm{I}_8} \pm \Gamma^5}{2},\quad
\mathcal{P}_{\pm} \equiv \frac{{\mathrm{I}_8} \pm {\Gamma^{y}}}{2}.
	\label{eq:definition_of_projective_operators}
}
Here, the coexistence of the two eigenvalues is ensured by the relation $\Gamma^5 {\Gamma^{y}} = {\Gamma^{y}} \Gamma^{5}$, which leads to $\mathcal{P}_{R/L} \mathcal{P}_{\pm} = \mathcal{P}_{\pm} \mathcal{P}_{R/L}$.
These projective operators fulfill the relations
\al{
\Gamma^A \mathcal{P}_{R/L} =
	\begin{cases}
		\mathcal{P}_{L/R} \Gamma^A & A = 0,1,2,3, \\
		\mathcal{P}_{R/L} \Gamma^A & A = y_1, y_2, 
	\end{cases}
\qquad
\Gamma^A \mathcal{P}_{\pm} =
	\begin{cases}
		\mathcal{P}_{\pm} \Gamma^A & A = 0,1,2,3, \\
		\mathcal{P}_{\mp} \Gamma^A & A = y_1, y_2, 
	\end{cases}
}
where we use them throughout our discussions on this manuscript.\footnote{
The following relations are useful to understand the formulas:
$\Gamma^\mu \Gamma^5 = - \Gamma^5 \Gamma^\mu$,
${\Gamma^{y_j}} \Gamma^5 = + \Gamma^5 \Gamma^{y_j}\,(j=1,2)$,
$\Gamma^\mu \Gamma^{y_j} = - \Gamma^{y_j} \Gamma^\mu$,
$\Gamma^\mu {\Gamma^{y}} = + {\Gamma^{y}} \Gamma^\mu$,
$\Gamma^{y_j} {\Gamma^{y}} = - {\Gamma^{y}} \Gamma^{y_j}$.
}
Now, it is easy to understand the decomposition of a 6d Dirac fermion $\Psi$,
\al{
\Psi = \Psi_{R+} + \Psi_{L+} + \Psi_{R-} + \Psi_{L-},
}
with the eigenstates of the two eigenvalues
\al{
\Psi_{R \pm} = \mathcal{P}_{R} \mathcal{P}_{\pm} \Psi,\quad
\Psi_{L \pm} = \mathcal{P}_{L} \mathcal{P}_{\pm} \Psi.
}

{
The corresponding projection operation for the 6d chirality should take the form
\al{
\mathcal{P}_{\Gamma^7 = \pm 1} \equiv \frac{I_8 \pm \Gamma^7}{2}.
}
The commutativities, $\mathcal{P}_{R/L} \mathcal{P}_{\Gamma^7 = \pm 1} = \mathcal{P}_{\Gamma^7 = \pm 1} \mathcal{P}_{R/L}$ and $\mathcal{P}_{\pm} \mathcal{P}_{\Gamma^7 = \pm 1} = \mathcal{P}_{\Gamma^7 = \pm 1} \mathcal{P}_{\pm}$ are easily understand by Eq.~(\ref{eq:Gamma_7}).
As we pointed out beforehand, the value of $\Gamma^7$ is determined when we fix the 4d chirality and the internal chirality, where the following correspondence is easily found [c.f. Eq.~(\ref{eq:6d_spinor_forms})]
\al{
\Psi_{R+} = \Psi_{R,\, \Gamma^7 = + 1},\quad
\Psi_{L+} = \Psi_{{L},\, \Gamma^7 = - 1},\quad
\Psi_{R-} = \Psi_{{R},\, \Gamma^7 = - 1},\quad
\Psi_{L-} = \Psi_{L,\, \Gamma^7 = + 1},
}
\vspace{-10mm}
\al{
\Psi_{\Gamma^7 = + 1} = \Psi_{R,\, \Gamma^7 = + 1} + \Psi_{L,\, \Gamma^7 = + 1},\quad
\Psi_{\Gamma^7 = - 1} = \Psi_{R,\, \Gamma^7 = - 1} + \Psi_{L,\, \Gamma^7 = - 1}.
}
The decomposition of the 6d Dirac fermion $\Psi$
\al{
\overline{\Psi} \Psi = 
	\overline{\Psi}_{\Gamma^7 = + 1} \Psi_{\Gamma^7 = - 1} +
	\overline{\Psi}_{\Gamma^7 = - 1} \Psi_{\Gamma^7 = + 1}
}
immediately leads to that the 6d bulk mass term vanishes when a 6d Weyl fermion is considered.
Being apparent from Eq.~(\ref{eq:Gamma_7}), two of the eigenvalues of $\Gamma^5$, $\Gamma^y$ and $\Gamma^7$ are independent, where {the other one} is automatically determined.
We adopt $\Gamma^5$ and $\Gamma^y$ as independent degrees of freedom, where this choice is {convenient} for the arguments developed in this manuscript.
}

For concrete decompositions of 6d spinor components, the following representation of the Gamma matrices is convenient,
\al{
\Gamma^\mu   &= {\mathrm{I}_2} \otimes \gamma^\mu =
	\begin{pmatrix} \gamma^\mu & 0 \\ 0 & \gamma^\mu  \end{pmatrix}, \notag \\
\Gamma^{y_1} &= i\sigma_1 \otimes \gamma^5 =
	\begin{pmatrix} 0 & i\gamma^5 \\ i\gamma^5 & 0 \end{pmatrix}, \notag \\
\Gamma^{y_2} &= i\sigma_2 \otimes \gamma^5 =
	\begin{pmatrix} 0 & \gamma^5 \\ -\gamma^5 & 0 \end{pmatrix},
}
where $\gamma^\mu$ represent the 4d part of the 6d Clifford algebra with the concrete forms
\al{
\gamma^\mu = \begin{pmatrix} 0 & \sigma^\mu \\ \overline{\sigma}^\mu & 0 \end{pmatrix}
\quad \text{with} \quad
\sigma^\mu = ({\mathrm{I}_2}, - {\boldsymbol \sigma}),\
\overline{\sigma}^\mu = ({\mathrm{I}_2}, {\boldsymbol \sigma}).
}
${\boldsymbol \sigma}$ means the set of the Pauli matrices $(\sigma_1, \sigma_2, \sigma_3)$.
$\gamma^5$ is defined in the same manner as
\al{
\gamma^5 = i \gamma^0 \gamma^1 \gamma^2 \gamma^3 =
	\begin{pmatrix} {\mathrm{I}_2} & 0 \\ 0 & -{\mathrm{I}_2} \end{pmatrix}.
}
The matrices representing the chiralities in 6d are expressed with the following diagonal forms
\al{
\Gamma^5 &= {\mathrm{I}_2} \otimes \gamma^5 =
	\begin{pmatrix} \gamma^5 & 0 \\ 0 & \gamma^5 \end{pmatrix}, \notag \\
\Gamma^{{y}} &= \sigma_3 \otimes {\mathrm{I}_4} =
	\begin{pmatrix} {\mathrm{1}_4} & 0 \\ 0 & -{\mathrm{I}_4} \end{pmatrix}, \notag \\
\Gamma^7 &= \sigma_3 \otimes \gamma^5 =
	\begin{pmatrix} \gamma^5 & 0 \\ 0 & -\gamma^5 \end{pmatrix}.
}
In this basis, the eigenstates of $\Gamma^5$ and {$\Gamma^{y}$} take the forms of
\al{
\Psi_{R+} = \begin{pmatrix} \xi_{R+} \\ 0 \\ 0 \\ 0 \end{pmatrix},\quad
\Psi_{L+} = \begin{pmatrix} 0 \\ \xi_{L+} \\ 0 \\ 0 \end{pmatrix},\quad
\Psi_{R-} = \begin{pmatrix} 0 \\ 0 \\ \xi_{R-} \\ 0 \end{pmatrix},\quad
\Psi_{L-} = \begin{pmatrix} 0 \\ 0 \\ 0 \\ \xi_{L-} \end{pmatrix},
	\label{eq:6d_spinor_forms}
}
where $\xi_{R\pm}$ and {$\xi_{L\pm}$} are two-component spinors.

Finally, let us look at the {4d subgroup of the} 6d Lorentz transformation of a 6d Dirac spinor $\Psi$, which is represented as
\al{
\Psi(x,y) \to \Psi'(x',y) = S({\omega}) \Psi(x,y),
}
with
\al{
S^{-1}({\omega}) \, \Gamma^\mu \, S({\omega}) = \Lambda^{\mu}_{\ \nu}{(\omega)} \, \Gamma^\nu.
	\label{eq:4d_Lorentz_transformation}
}
$S({\omega})$ is the 4d Lorentz transformation matrix for 6d spinors and is expressed with the boost/rotation parameters $\omega_{\mu\nu}$ in the present representation of $\Gamma^A$ as
\al{
S({\omega}) &= \exp\left(- \frac{i}{2} \omega_{\mu\nu} J^{\mu\nu}\right)
\quad \text{with} \quad J^{\mu\nu} \equiv \frac{i}{4} \left[ \Gamma^\mu, \Gamma^\nu \right]
	= \frac{i}{4} \begin{pmatrix} \left[ \gamma^\mu, \gamma^\nu \right] & 0 \\ 0 & \left[ \gamma^\mu, \gamma^\nu \right]\end{pmatrix} \notag \\
&=
\begin{pmatrix}
S_R({\omega}) & 0 & 0 & 0 \\
0 & S_L({\omega}) & 0 & 0 \\
0 & 0 & S_R({\omega}) & 0 \\
0 & 0 & 0 & S_L({\omega})
\end{pmatrix},
	\label{eq:form_of_SLorentz}
}
where $S_{R/L}({\omega})$ are the representations for 4d two-component spinors with right/left chiralities defined as
\al{
S_R({\omega}) &= \exp\left({\frac{1}{8} \omega_{\mu\nu} \left( \sigma^\mu \overline{\sigma}^\nu - \sigma^\nu \overline{\sigma}^\mu \right)}\right), \notag \\
S_L({\omega}) &= \exp\left({\frac{1}{8} \omega_{\mu\nu} \left( \overline{\sigma}^\mu \sigma^\nu - \overline{\sigma}^\nu \sigma^\mu \right)}\right). 
}
Here, the form in Eq.~(\ref{eq:4d_Lorentz_transformation}) is converted to the two-component sense of
\al{
S_R^{-1}({\omega}) \, \sigma^\mu S_L({\omega}) = \Lambda^{\mu}_{\ \nu}{(\omega)} \, \sigma^\nu,\quad
S_L^{-1}({\omega}) \, \overline{\sigma}^\mu S_R({\omega}) = \Lambda^{\mu}_{\ \nu}{(\omega)} \, \overline{\sigma}^\nu.
}
As we can clearly understand from Eq.~(\ref{eq:form_of_SLorentz}), right- and left-handed components do not mix each other under the 4d Lorentz transformation.

\subsection{Requirement via Variational Principle}

The variational principal is the powerful doctrine for determining the forms of equation of motions and boundary conditions. 
Let us take a variation of $\Psi(x,y)$ in the action in Eq.~(\ref{eq:6d_action})
\al{
\delta S = \int \! d^{4} x \int_{0}^{L_{1}} \!\!\!\! dy_{1} \int_{0}^{L_{2}} \!\!\!\! dy_{2} \,
	\overline{\Psi}(x,y) \left( i \Gamma^A \del_A - M \right) \delta\Psi(x,y),
}
which, after integrations by parts, results in
\al{
\delta S &= \int \! d^{4} x \int_{0}^{L_{1}} \!\!\!\! dy_{1} \int_{0}^{L_{2}} \!\!\!\! dy_{2} \,
	\overline{\Psi}(x,y) \left( -i \Gamma^A \overset{\longleftarrow}{\del_A} - M \right) \delta\Psi(x,y) \notag \\
&+ \int \! d^{4} x \Bigg\{ \int_{0}^{L_{2}} \!\!\!\! dy_{2}
	\left[ \overline{\Psi}(x,y) \, i \Gamma^{y_1}  \delta\Psi(x,y) \right]_{y_1 = 0}^{{y_1 = L_1}} +
	{\int_{0}^{L_{1}} \!\!\!\! dy_{1}}
	\left[ \overline{\Psi}(x,y) \, i \Gamma^{y_2}  \delta\Psi(x,y) \right]_{y_2 = 0}^{{y_2 = L_2}} \Bigg\}.
}
The above form gives us the EOM of $\overline{\Psi}$ (equivalently for $\Psi$)
\al{
\overline{\Psi}(x,y) \left( -i \Gamma^A \overset{\longleftarrow}{\del_A} - M \right) = 0,\quad
\left(\text{or } \left( i \Gamma^A \del_A - M \right) \Psi(x,y) = 0 \right){.}
	\label{eq:EOM_for_Psi}
}
{Also, focusing on the above form leads to an interpretation of vanishing the surface terms for consistency}\footnote{
See~\cite{Gabriel:2004ua} for a discussion on BCs in 6d based on the variational principle.
}
\al{
{\Big[ \overline{\Psi}(x,y) \Gamma^{y_1}  \delta\Psi(x,y)\Big]_{y_1 = 0, L_1}} = 0  \quad\text{and}\quad
{\Big[ \overline{\Psi}(x,y) \Gamma^{y_2}  \delta\Psi(x,y)\Big]_{y_2 = 0, L_2}} = 0.
	\label{eq:original_form}
}
{{Here, $\delta \Psi(x,y)$ is an arbitrary variation of $\Psi(x,y)$, and $\delta \Psi(x,y)$ can be independent of $\Psi(x,y)$.}
It would seem that these forms are not so friendly for discussing general features of BCs.
Thereby at first, we focus on the following forms}
\al{
{\Big[ \overline{\Psi}(x,y) \Gamma^{y_1}  \Psi(x,y)\Big]_{y_1 = 0, L_1}} = 0  \quad\text{and}\quad
{\Big[ \overline{\Psi}(x,y) \Gamma^{y_2}  \Psi(x,y)\Big]_{y_2 = 0, L_2}} = 0,
	\label{eq:current_form}
}
{which {look} a necessary condition for (\ref{eq:original_form}).
{In general}, it is nontrivial whether the two relations in Eq.~(\ref{eq:original_form}) are satisfied when $\Psi(x,y)$  [{\it not} $\delta \Psi(x,y)$] takes a configuration derived from Eq.~(\ref{eq:current_form}) at the boundaries.
Later, we check that all of the configurations of $\Psi(x,y)$ via Eq.~(\ref{eq:current_form}) result in the realization of (\ref{eq:original_form}).
When we start from (\ref{eq:current_form}), every possible BC should fulfill the two requirements in Eq.~(\ref{eq:current_form}).}

\subsection{Classification of Boundary Conditions along $y_1$ {Direction}
\label{sec:bc_y1}}

As discussed, e.g. in Refs.~\cite{Fujimoto:2011kf}, the classification of boundary conditions arrives at analyzing the current forms which appear as surface terms in the variation of actions.
Now, two extra spacial directions are {in existence}, and then we should analyze the two types of the current form shown in Eq.~(\ref{eq:current_form}) individually.

Firstly in this section, we focus on the former form in Eq.~(\ref{eq:current_form}) for the $y_1$ direction.
By use of the following relations (which are easily derived from Eqs.~(\ref{eq:Gamma_Clifford}) and (\ref{eq:definition_of_projective_operators})),
\al{
\Gamma^0 \Gamma^{y_j} \Psi_{R/L \pm} &= \mathcal{P}_{L/R} \mathcal{P}_\mp \left( \Gamma^0 \Gamma^{y_j} \Psi_{R/L \pm} \right)\quad (j=1,2), \notag \\
\left( \Psi_{R/L \pm} \right)^\dagger &= \left( \Psi_{R/L \pm} \right)^\dagger \mathcal{P}_{R/L} \mathcal{P}_{\pm},
	\label{eq:projectors_relation_RLandpm}
}
we find the following transformation
\al{
0 &= {\Big[ \overline{\Psi}(x,y)   \Gamma^{y_1}  \Psi(x,y) \Big]_{y_1 = 0, L_1}} \notag \\
&= \left[ \rho_R^\dagger \lambda_R + \lambda_R^\dagger \rho_R \right]_{y_1 = 0, L_1},
	\label{eq:current_condition_y1}
}
with
\al{
\rho_R = \begin{pmatrix} \Psi_{R+} \\ \Psi_{R-} \end{pmatrix},\quad
\lambda_R = \begin{pmatrix} \Gamma^0 \Gamma^{y_1} \Psi_{L-} \\ \Gamma^0 \Gamma^{y_1} \Psi_{L+} \end{pmatrix}.
}

Interestingly, under the realization of the condition in Eq.~(\ref{eq:current_condition_y1}), the final form of Eq.~(\ref{eq:current_condition_y1}) is reformulated with a nonzero real parameter $c_0$ as
\al{
\left| \rho_R + c_0 \lambda_R \right|^2 = \left| \rho_R - c_0 \lambda_R \right|^2
	\quad \text{at} \ y_1 =0, L_1.
	\label{eq:extended_current_form_y1}
}
Because the nonzero components of $\Psi_{R\pm}$ and $\Psi_{L\pm}$ are represented by two-component spinors, a general solution to (\ref{eq:extended_current_form_y1}) turns out to be given by
\al{
\rho_R + c_0 \lambda_R = U \left( \rho_R - c_0 \lambda_R \right)
	\quad \text{at} \ y_1 =0, L_1,
}
with $U \in U(4)$.
After some straightforward calculations, we reach the following form
\al{
\left( {\mathrm{I}_4} - U \right) \begin{pmatrix} \xi_{R+} \\ \xi_{R-} \end{pmatrix}
=
\left( {\mathrm{I}_4} + U \right) i c_0 \begin{pmatrix} \xi_{L-} \\ \xi_{L+} \end{pmatrix}
	\quad \text{at} \ y_1 =0, L_1.
	\label{eq:U(4)_solution_y1}
}
Here, to maintain the 4d Lorentz structure shown in Eq.~(\ref{eq:form_of_SLorentz}), where each {right-handed} and {left-handed} components are 4d-Lorentz transformed individually, $U$ should not contain spinor indices.
Accordance with this requirement leads to the reduction of the form in Eq.~(\ref{eq:U(4)_solution_y1}) into
\al{
\left( {\mathrm{I}_4} - \mathcal{U} \otimes {\mathrm{I}_2} \right) \begin{pmatrix} \xi_{R+} \\ \xi_{R-} \end{pmatrix}
=
\left( {\mathrm{I}_4} + \mathcal{U} \otimes {\mathrm{I}_2} \right) i c_0 \begin{pmatrix} \xi_{L-} \\ \xi_{L+} \end{pmatrix}
	\quad \text{at} \ y_1 =0, L_1,
	\label{eq:U(2)_solution_y1}
}
where $\mathcal{U}$ is a two-by-two unitary matrix belonging to the $U(2)$ group, which acts only on the ``flavor" DOFs ($+$ or $-$).

In the following part, we classify possible BCs which preserve the 4d Lorentz structure from the form in Eq.~(\ref{eq:U(2)_solution_y1}).
After a glance at (\ref{eq:U(2)_solution_y1}), we recognize that the right-handed spinors are mixed with the left-handed ones in general, which {could violate} the 4d Lorentz invariance.
Meaningful solutions are a subset of general solutions, where {right-handed} and {left-handed} components are 4d-Lorentz transformed separately.
Under the necessity, the following three classes are possible,
\al{
&{\text{Type }} \text{\GkI-}{y_1} \ \ :\quad
	\mathcal{U} = +{\mathrm{I}_2}, \\
&{\text{Type }} \text{{\GkII}-}{y_1} \ :\quad
	\mathcal{U} = -{\mathrm{I}_2}, \\
&{\text{Type }} \text{{\GkIII}-}{y_1} :\quad
	\mathcal{U} = \vec{n} \cdot \vec{\sigma} =
	\begin{pmatrix}
	\cos{\theta} & e^{{-i\phi}} \sin{\theta} \\
	e^{{i\phi}} \sin{\theta} & - \cos{\theta}
	\end{pmatrix},
}
with $\vec{\sigma} = (\sigma_1, \sigma_2, \sigma_3)$ and $\vec{n} = {(\cos{\phi} \sin{\theta}, \sin{\phi} \sin{\theta}, \cos{\theta})}$, which is the three-dimensional vector specifying a point of a unit two-dimensional sphere $S^2$.\footnote{
{A more detailed description is found in the separate publication~\cite{Fujimoto:2016rbr} on the classification.}
}
Here, the possibility $\mathcal{U} = -\vec{n} \cdot \vec{\sigma}$ gives us no new information after the consideration of $\mathcal{U} = +\vec{n} \cdot \vec{\sigma}$.
Therefore, we dropped it.
Each of the conditions is rewritten as the projections of a part of the 6d spinor components as
\al{
&{\text{Type }} \text{\GkI-}{y_1} \ \ :\quad
	\begin{pmatrix} \xi_{L-}(x,y) \\ \xi_{L+}(x,y) \end{pmatrix} = 0
	\quad {\text{at} \ y_1 = 0, L_1},
	\label{eq:type-I_condition_y1} \\
&{\text{Type }} \text{{\GkII}-}{y_1} \ :\quad
	\begin{pmatrix} \xi_{R+}(x,y) \\ \xi_{R-}(x,y) \end{pmatrix} = 0
	\quad {\text{at} \ y_1 = 0, L_1},
	\label{eq:type-II_condition_y1} \\
&{\text{Type }} \text{{\GkIII}-}{y_1} :\quad
	\mathcal{P'}_{\vec{n}\cdot\vec{\sigma}=-1}(\phi,\theta) \begin{pmatrix} \xi_{R+}(x,y) \\ \xi_{R-}(x,y) \end{pmatrix} = 0
	\quad \text{and} \quad
	\mathcal{P'}_{\vec{n}\cdot\vec{\sigma}=+1}(\phi,\theta) \begin{pmatrix} \xi_{L-}(x,y) \\ \xi_{L+}(x,y) \end{pmatrix} = 0
	\quad {\text{at} \ y_1 = 0, L_1}.
	\label{eq:type-III_condition_y1}
}
The projectors for pairs of two-component spinors $\mathcal{P'}_{\vec{n}\cdot\vec{\sigma}=\pm 1}(\phi,\theta)$ pick up the eigenstates of the variable $\vec{n}\cdot\vec{\sigma}$ being $+1$ or $-1$, which is well defined because of $(\vec{n}\cdot\vec{\sigma})^2 = {\mathrm{I}_2}$ (where we used $\vec{n} \cdot \vec{n} = 1$), $(\vec{n}\cdot\vec{\sigma})^\dagger = \vec{n}\cdot\vec{\sigma}$, as
\al{
\mathcal{P'}_{\vec{n}\cdot\vec{\sigma}=\pm 1}(\phi,\theta) = \left(\frac{{\mathrm{I}_2} \pm \vec{n}\cdot\vec{\sigma}}{2}\right) \otimes {\mathrm{I}_2}.
}
The two conditions in Eq.~(\ref{eq:type-III_condition_y1}) for {Type \GkIII}-$y_1$ are unified in a single form as
\al{
{
\widetilde{\Gamma} \begin{pmatrix} \xi_{R+} \\ \xi_{R-} \\ \xi_{L-} \\ \xi_{L+} \end{pmatrix}
= + \begin{pmatrix} \xi_{R+} \\ \xi_{R-} \\ \xi_{L-} \\ \xi_{L+} \end{pmatrix}
}
\quad {\text{at} \ y_1 = 0, L_1},
\quad \text{with} \
\widetilde{\Gamma} \equiv
\begin{pmatrix}
(\vec{n}\cdot\vec{\sigma}) \otimes {\mathrm{I}_2} & 0 \\
0 & -(\vec{n}\cdot\vec{\sigma}) \otimes {\mathrm{I}_2}
\end{pmatrix},
}
which is equivalent to {the form at $y_1 = 0, L_1$,}
\al{
{(\vec{n} \cdot \vec{\Sigma})
\begin{pmatrix}
\xi_{R+} \\ \xi_{L+} \\ \xi_{R-} \\ \xi_{L-}
\end{pmatrix}
=
\begin{pmatrix}
\xi_{R+} \\ \xi_{L+} \\ \xi_{R-} \\ \xi_{L-}
\end{pmatrix}}, \quad
\text{with} \
\vec{n} \cdot \vec{\Sigma}
\equiv
\begin{pmatrix}
\cos{\theta} \,{\mathrm{I}_2} & 0 & e^{-i\phi} \sin{\theta} \,{\mathrm{I}_2} & 0 \\
0 & \cos{\theta} \,{\mathrm{I}_2} & 0 & -e^{i\phi} \sin{\theta} \,{\mathrm{I}_2} \\
e^{i\phi} \sin{\theta} \,{\mathrm{I}_2} & 0 & -\cos{\theta} \,{\mathrm{I}_2} & 0 \\
0 & -e^{-i\phi} \sin{\theta} \,{\mathrm{I}_2} & 0 & -\cos{\theta} \,{\mathrm{I}_2}
\end{pmatrix},
	\label{eq:6d_type-III_y1}
}
where the eight-by-eight flavor matrices $\vec{\Sigma} = (\Sigma_1, \Sigma_2, \Sigma_3)$ are defined as
\al{
\Sigma_1 = -i \Gamma^{y_1},\quad
\Sigma_2 = -i \Gamma^{y_2} \Gamma^5,\quad
\Sigma_3 = {\Gamma^{y}},
}
which are found to satisfy the relations
\al{
\{ \Sigma_k, \Sigma_l \} &= 2 \delta_{kl} \, {\mathrm{I}_8},\quad
(\Sigma_k)^\dagger = \Sigma_k \quad (k,l = 1,2,3),
	\label{eq:Sigma_relations} \\
{(\Gamma^0 \Gamma^{y_1}) \Sigma_k} &= {- \Sigma_k (\Gamma^0 \Gamma^{y_1})},
	\label{eq:Sigma_anticommutation}
}
With the property $(\vec{n} \cdot \vec{\Sigma})^2 = {\mathrm{I}_8}$ (which is easily derived from Eq.~(\ref{eq:Sigma_relations})),
the following projectors for eight-component spinors are consistently defined as
\al{
\mathcal{P}_{\vec{n}\cdot\vec{\Sigma}=\pm 1}(\phi,\theta) = \left(\frac{{\mathrm{I}_8} \pm \vec{n}\cdot\vec{\Sigma}}{2}\right).
}
{The properties also hold for the projectors [easily derived with (\ref{eq:Sigma_relations}) and (\ref{eq:Sigma_anticommutation})],
\al{
\Gamma^0 \Gamma^{y_1} \Psi_{\vec{n}\cdot\vec{\Sigma}=\pm 1} &= \mathcal{P}_{\vec{n}\cdot\vec{\Sigma}=\mp 1}(\phi, \theta) \left( \Gamma^0 \Gamma^{y_1} \Psi_{\vec{n}\cdot\vec{\Sigma}=\pm 1} \right), \notag \\
\left( \Psi_{\vec{n}\cdot\vec{\Sigma}=\pm 1} \right)^\dagger &= \left( \Psi_{\vec{n}\cdot\vec{\Sigma}=\pm 1} \right)^\dagger \mathcal{P}_{\vec{n}\cdot\vec{\Sigma}=\pm 1}(\phi, \theta),
	\label{eq:projectors_relation_ndotSigma}
}
where the counterpart for $\mathcal{P}_{L/R}$ and $\mathcal{P}_{\pm}$ are found in Eq.~(\ref{eq:projectors_relation_RLandpm}).}
{$\Psi_{\vec{n}\cdot\vec{\Sigma}=\pm 1}$ are the eigenstates of $\vec{n}\cdot\vec{\Sigma}$, which are straightforwardly defined as $\Psi_{\vec{n}\cdot\vec{\Sigma}=\pm 1} \equiv \mathcal{P}_{\vec{n}\cdot\vec{\Sigma}=\pm 1}(\phi, \theta) \Psi$.}

Now, the three possible BCs in Eqs.~(\ref{eq:type-I_condition_y1})--(\ref{eq:type-III_condition_y1}) are represented as single projections for a 6d Dirac spinor {by}
\al{
&{\text{Type }} \text{\GkI-}{y_1} \ \ :\quad
	\mathcal{P}_L \Psi(x,y) = 0  & &\text{at} \ y_1 = 0,L_1,
	\label{eq:type-I_condition_6d_y1} \\
&{\text{Type }} \text{{\GkII}-}{y_1} \ :\quad
	\mathcal{P}_R \Psi(x,y) = 0  & &\text{at} \ y_1 = 0,L_1,
	\label{eq:type-II_condition_6d_y1} \\
&{\text{Type }} \text{{\GkIII}-}{y_1} :\quad
	\mathcal{P}_{\vec{n}\cdot\vec{\Sigma} = -1}(\phi,\theta) \Psi(x,y) = 0 & &\text{at} \ y_1 = 0,L_1.
	\label{eq:type-III_condition_6d_y1}
}
We note that the eigenstate of $\vec{n}\cdot\vec{\Sigma} = -1$ is a mixture of two types of eigenvalues, $R/L$ and $\pm$ in general, but we can find two exceptional points of $(\phi,\theta) = (0, \pi)$ and $(0,0)$, where $\mathcal{P}_{\vec{n}\cdot\vec{\Sigma} = -1}(\phi,\theta)$ becomes equivalent to $\mathcal{P}_{\pm}$ as
\al{
&{\text{Type }} \text{{\GkIII}-}{y_1} \ \text{when} \ (\phi, \theta) = (0, \pi) : \quad
	\mathcal{P}_{+} \Psi(x,y) = 0 \qquad \text{at} \ y_1 = 0,L_1,
	\label{eq:type-plus_condition_6d_y1} \\
&{\text{Type }} \text{{\GkIII}-}{y_1} \ \text{when} \ (\phi, \theta) = (0, 0) \, : \quad
	\mathcal{P}_{-} \Psi(x,y) = 0 \qquad \text{at} \ y_1 = 0,L_1.
	\label{eq:type-minus_condition_6d_y1}
}

{Finally, we back to the original form in Eq.~(\ref{eq:original_form}) of the form.
By use of the properties of (\ref{eq:projectors_relation_RLandpm}) and (\ref{eq:projectors_relation_ndotSigma}),
The following transformations are possible at $y_1 = 0, L_1$,
\al{
\overline{\Psi} \Gamma^{y_1} \delta \Psi = \Psi^\dagger \Gamma^0 \Gamma^{y_1} \delta \Psi 
&= \Psi^\dagger_{R} \Gamma^0 \Gamma^{y_1} \delta \Psi_{L} + \Psi^\dagger_{L} \Gamma^0 \Gamma^{y_1} \delta \Psi_{R} \notag \\
&= \Psi^\dagger_{\vec{n}\cdot\vec{\Sigma}= +1} \Gamma^0 \Gamma^{y_1} \delta \Psi_{\vec{n}\cdot\vec{\Sigma}= -1} + \Psi^\dagger_{\vec{n}\cdot\vec{\Sigma}= -1} \Gamma^0 \Gamma^{y_1} \delta \Psi_{\vec{n}\cdot\vec{\Sigma}= +1}.
	\label{eq:variation_condition_check_y1}
}
Now, {under the assumption that $\Psi(x,y)$ and $\delta \Psi(x,y)$ obey the same conditions at the boundaries,} it is obvious that when $\Psi(x,y)$ takes the values designated in either of (\ref{eq:type-I_condition_6d_y1}), (\ref{eq:type-II_condition_6d_y1}) or (\ref{eq:type-III_condition_6d_y1}), the original condition for the BC along the $y_1$ directions is satisfied even though the corresponding unconstrained part is still arbitrary.}

\subsection{Classification of Boundary Conditions along $y_2$ {Direction}
\label{sec:bc_y2}}

The discussion for the latter form in Eq.~(\ref{eq:current_form}) for the $y_2$ direction is completely parallel to the one made in the previous section~\ref{sec:bc_y1}, and then we focus only on a major part of discussions without mentioning details.

The corresponding current form is transformed as
\al{
0 &= {\Big[ \overline{\Psi}(x,y) \Gamma^{y_2}  \Psi(x,y)\Big]_{y_2 = 0, L_2}} \notag \\
&= \left[ {\rho'_R}^{\dagger} \lambda'_R + {\lambda'_R}^{\dagger} \rho'_R \right]_{y_2 = 0, L_2},
	\label{eq:current_condition_y2}
}
with
\al{
\rho'_R = \begin{pmatrix} \Psi_{R+} \\ \Psi_{R-} \end{pmatrix},\quad
\lambda'_R = \begin{pmatrix} \Gamma^0 \Gamma^{y_2} \Psi_{L-} \\ \Gamma^0 \Gamma^{y_2} \Psi_{L+} \end{pmatrix}.
}
The consistent solutions are derived from the following corresponding condition (of Eq.~(\ref{eq:U(2)_solution_y1})) with a nonzero real constant $c'_0$
\al{
\left( {\mathrm{I}_4} - \mathcal{U} \otimes {\mathrm{I}_2} \right) \begin{pmatrix} \xi_{R+} \\ \xi_{R-} \end{pmatrix}
=
\left( {\mathrm{I}_4} + \mathcal{U} \otimes {\mathrm{I}_2} \right) (- c'_0) \begin{pmatrix} -\xi_{L-} \\ \xi_{L+} \end{pmatrix}
	\quad \text{at} \ {y_2} = 0, {L_2,}
	\label{eq:U(2)_solution_y2}
}
with requesting the individual 4d Lorentz transformations shown in Eq.~(\ref{eq:form_of_SLorentz}).
Here {apparently}, the conditions for $\mathcal{U}$ basically take the same forms as
\al{
&{\text{Type }} \text{\GkI-}{y_2} \ \ :\quad
	\mathcal{U} = +{\mathrm{I}_2}, \\
&{\text{Type }} \text{\GkII-}{y_2} \ :\quad
	\mathcal{U} = -{\mathrm{I}_2}, \\
&{\text{Type }} \text{\GkIII-}{y_2} :\quad
	\mathcal{U} = \vec{n'} \cdot \vec{\sigma} =
	\begin{pmatrix}
	\cos{\theta'} & e^{{-i\phi'}} \sin{\theta'} \\
	e^{{i\phi'}} \sin{\theta'} & - \cos{\theta'}
	\end{pmatrix},
}
where in the case of {Type \GkIII}-$y_2$, we can take different variables $\phi'$ and $\theta'$ for parameterizing a position on  a unit two-dimensional sphere $S^2$.
On the other hand, the corresponding form of Eq.~(\ref{eq:6d_type-III_y1}) takes the different shape {(at $y_2 = 0, L_2$)} as
\al{
{(\vec{n'} \cdot \vec{\Sigma'})
\begin{pmatrix}
\xi_{R+} \\ \xi_{L+} \\ \xi_{R-} \\ \xi_{L-}
\end{pmatrix}
=
\begin{pmatrix}
\xi_{R+} \\ \xi_{L+} \\ \xi_{R-} \\ \xi_{L-}
\end{pmatrix}},
\quad \text{with} \ 
\vec{n'} \cdot \vec{\Sigma'}
\equiv
\begin{pmatrix}
\cos{\theta'} \,{\mathrm{I}_2} & 0 & e^{-i\phi'} \sin{\theta'} \,{\mathrm{I}_2} & 0 \\
0 & \cos{\theta'} \,{\mathrm{I}_2} & 0 & e^{i\phi'} \sin{\theta'} \,{\mathrm{I}_2} \\
e^{i\phi'} \sin{\theta'} \,{\mathrm{I}_2} & 0 & -\cos{\theta'} \,{\mathrm{I}_2} & 0 \\
0 & e^{-i\phi'} \sin{\theta'} \,{\mathrm{I}_2} & 0 & -\cos{\theta'} \,{\mathrm{I}_2}
\end{pmatrix},
	\label{eq:6d_type-III_y2}
}
depicted with the different set of the eight-by-eight flavor matrices $\vec{\Sigma'} = (\Sigma'_1, \Sigma'_2, \Sigma'_3)$ defined as
\al{
\Sigma'_1 = -i \Gamma^{y_1} \Gamma^5,\quad
\Sigma'_2 = -i \Gamma^{y_2},\quad
\Sigma'_3 = {\Gamma^{y}}.
}

By use of the similar projective matrices
\al{
\mathcal{P}_{\vec{n'}\cdot\vec{\Sigma'}=\pm 1}(\phi',\theta') = \left(\frac{{\mathrm{I}_8} \pm \vec{n'}\cdot\vec{\Sigma'}}{2}\right),
}
the three possible BCs are represented in terms of {the} 6d Dirac fermion as follows,
\al{
&{\text{Type }} \text{\GkI-}{y_2} \ \ :\quad
	\mathcal{P}_L \Psi(x,y) = 0  & &\text{at} \ y_2 = 0,L_2,
	\label{eq:type-I_condition_6d_y2} \\
&{\text{Type }} \text{{\GkII}-}{y_2} \ :\quad
	\mathcal{P}_R \Psi(x,y) = 0  & &\text{at} \ y_2 = 0,L_2,
	\label{eq:type-II_condition_6d_y2} \\
&{\text{Type }} \text{{\GkIII}-}{y_2} :\quad
	\mathcal{P}_{\vec{n'}\cdot\vec{\Sigma'} = -1}(\phi',\theta') \Psi(x,y) = 0 & &\text{at} \ y_2 = 0,L_2.
	\label{eq:type-III_condition_6d_y2}
}
The following specific cases also exist for {Type \GkIII}-$y_2$,
\al{
&{\text{Type }} \text{{\GkIII}-}{y_2} \ \text{when} \ (\phi', \theta') = (0, \pi) : \quad
	\mathcal{P}_{+} \Psi(x,y) = 0 \qquad \text{at} \ y_2 = 0,L_2,
	\label{eq:type-plus_condition_6d_y2} \\
&{\text{Type }} \text{{\GkIII}-}{y_2} \ \text{when} \ (\phi', \theta') = (0, 0) \, : \quad
	\mathcal{P}_{-} \Psi(x,y) = 0 \qquad \text{at} \ y_2 = 0,L_2.
	\label{eq:type-minus_condition_6d_y2}
}
{The properties shown in Eqs.~(\ref{eq:Sigma_relations}), (\ref{eq:Sigma_anticommutation}) and (\ref{eq:projectors_relation_ndotSigma}) are maintained after the replacements, $\Sigma_k \to \Sigma'_k$ and $\Gamma^{y_1} \to \Gamma^{y_2}$.
Thereby, the discussion around Eq.~(\ref{eq:variation_condition_check_y1}) is applicable also for the $y_2$ direction.}

\subsection{Comment on 6d Weyl case
\label{sec:bc_Weyl}}

{Here, we briefly comment on the possible BCs when the 6d fermion is Weyl-type, whose 6d chirality is defined in Eq.~(\ref{eq:Gamma_7}).
In the Weyl case, possible BCs are classified in the same way,
but, no $U(2)$ rotation is possible, which is found in Eq.~(\ref{eq:U(2)_solution_y1}).
Thereby, only the following types are realizable along $y_i\,(i=1,2)$ for $\Psi_{\Gamma^7 = + 1}$ or $\Psi_{\Gamma^7 = - 1}$,}
\al{
&{\text{Type }} \text{\GkI-}{y_i} \ \ :\quad
	\mathcal{P}_L \Psi_{\Gamma^7 = \pm 1}(x,y) = 0  & &\text{at} \ y_i = 0,L_i,
	\label{eq:type-I_condition_6d_Weyl} \\
&{\text{Type }} \text{{\GkII}-}{y_i} \ :\quad
	\mathcal{P}_R \Psi_{\Gamma^7 = \pm 1}(x,y) = 0  & &\text{at} \ y_i = 0,L_i.
	\label{eq:type-II_condition_6d_Weyl}
}

{We add a few sentences on Type-$\text{\GkIII}$.
From Eqs.~(\ref{eq:6d_type-III_y1}) [and/or (\ref{eq:6d_type-III_y2})], we recognize that $\vec{n}^{(')} \cdot \vec{\Sigma}^{(')}$ does not {become} $\Gamma^5$ and $\Gamma^7$ in any choice of $(\phi^{(')},\theta^{(')})$.
Thereby, the following condition is not derived in the present framework,}
\al{
{\mathcal{P}_{\Gamma^7 = {\pm 1}} \Psi_{\Gamma^7 = \pm 1}(x,y) = 0}  & &{\text{at} \ y_i = 0,L_i.}
}
{In Type-$\text{\GkIII}$ for a 6d Weyl fermion, only the cases $(\phi^{(')},\theta^{(')}) = (0,\pi)$ or $(0,0)$ are meaningful.
But, they are equivalent to Type-I or Type-$\text{\GkII}$ for a 6d Weyl fermion, which means that no new possibility is induced.}

\section{KK Expansion in Two Specific BCs
\label{sec:KK_expansion}}

As we concluded in the previous section~\ref{sec:setup_and_bcs}, for each of the $y_1$ and $y_2$ directions,
the three types of BCs are possible (Eqs.~(\ref{eq:type-I_condition_6d_y1})--(\ref{eq:type-III_condition_6d_y1}) for $y_1$, Eqs.~(\ref{eq:type-I_condition_6d_y2})--(\ref{eq:type-III_condition_6d_y2}) for $y_2$, respectively).

The cases with phenomenological interests are the following two ones,
\al{
&{\text{Case \GkI \ [Type \GkI-$y_1$ and Type \GkI-$y_2$]}} \ \ \, :\quad \Psi_{L\pm}(x,y) = 0
	\quad \text{at}\ y_1 = 0, L_1 \ {\text{and}} \ y_2 = 0, L_2, 
	\label{eq:BC_case-I} \\
&{\text{Case \GkII \ [Type \GkII-$y_1$ and Type \GkII-$y_2$]}} \hspace{0.5pt} :\quad \Psi_{R\pm}(x,y) = 0
	\quad \text{at}\ y_1 = 0, L_1 \ {\text{and}} \ y_2 = 0, L_2,
	\label{eq:BC_case-II}
}
where left-handed components in Eq.~(\ref{eq:BC_case-I}) or right-handed components in Eq.~(\ref{eq:BC_case-II}) are set to be zero at the boundaries and no corresponding zero mode is expected.
On the other hand, the components with opposite 4d {chiralities} [right modes in Eq.~(\ref{eq:BC_case-I}) and left modes in Eq.~(\ref{eq:BC_case-II})] have no restriction from the BCs and corresponding chiral modes can be realized.

When we adopt {Type-\GkIII} BCs, the lowest part may not be chiral since {Type \GkIII} does not contain the chiral cases {(Type I and Type \GkII)}, and then this possibility has less interests in the phenomenological point of view.
On the other hand, we find some interesting aspects in the theoretical point of view, and briefly investigate this class by focusing on one of the simplest cases of
\al{
&{\text{Case \GkIII \ [Type \GkIII-(\ref{eq:type-plus_condition_6d_y1}) and Type \GkIII-(\ref{eq:type-plus_condition_6d_y2})]}} :\quad \Psi_{L/R +}(x,y) = 0
	\quad \text{at}\ y_1 = 0, L_1 \ {\text{and}} \ y_2 = 0, L_2, 
	\label{eq:BC_case-III}
}
with the specific choice of the parameters on $S^2$, $(\phi,\theta) = (\phi',\theta') = (0, \pi)$.

\subsection{Case {\GkII} --- a Chiral Possibility
\label{sec:case-II}}

At first, we focus on the case where the emergence of left-handed 4d chiral modes are expected.
Before concrete discussion on this case, let us comment on {Case \GkI} where right-handed {chiral} 4d modes would occur.
All the following discussions are parallel to those of {Case I} (which are expected) and we give a brief note on {Case \GkI} after the end of the discussions on {Case \GkII}.

The EOM of the 6d fermion $\Psi$ shown in Eq.~(\ref{eq:EOM_for_Psi}) is expressed in a more concrete manner,
\al{
\left[ i \Gamma^\mu \del_\mu + i \Gamma^{y_1} \del_{y_1} + i \Gamma^{y_2} \del_{y_2} -M \right] \Psi(x,y) = 0,
	\label{eq:6d_Dirac_eq_decomponsed}
}
which can be decomposed into the following subset by use of the two types of the projective matrices in Eq.~(\ref{eq:definition_of_projective_operators}) as
\al{
i \Gamma^\mu \del_\mu \Psi_{L+} + \left( i \Gamma^{y_1} \del_{y_1} + i \Gamma^{y_2} \del_{y_2} \right) \Psi_{R-}
	- M \Psi_{R+} &= 0,
	\label{eq:6d_Dirac_eq__R+} \\
i \Gamma^\mu \del_\mu \Psi_{L-} + \left( i \Gamma^{y_1} \del_{y_1} + i \Gamma^{y_2} \del_{y_2} \right) \Psi_{R+}
	- M \Psi_{R-} &= 0,
	\label{eq:6d_Dirac_eq__R-} \\
i \Gamma^\mu \del_\mu \Psi_{R+} + \left( i \Gamma^{y_1} \del_{y_1} + i \Gamma^{y_2} \del_{y_2} \right) \Psi_{L-}
	- M \Psi_{L+} &= 0,
	\label{eq:6d_Dirac_eq__L+} \\
i \Gamma^\mu \del_\mu \Psi_{R-} + \left( i \Gamma^{y_1} \del_{y_1} + i \Gamma^{y_2} \del_{y_2} \right) \Psi_{L+}
	- M \Psi_{L-} &= 0,
	\label{eq:6d_Dirac_eq__L-}
}
by casting the following products of the projectors from the left-hand sizes of (\ref{eq:6d_Dirac_eq_decomponsed}),
$\mathcal{P}_{R} \mathcal{P}_{+}$,
$\mathcal{P}_{R} \mathcal{P}_{-}$,
$\mathcal{P}_{L} \mathcal{P}_{+}$,
$\mathcal{P}_{L} \mathcal{P}_{-}$, respectively.
These forms can be written by use of Eqs.~(\ref{eq:Gamma_sq_relations}) and (\ref{eq:eigenstate_Gammay3}) as
\al{
i \Gamma^\mu \del_\mu \Psi_{L\pm} + \left( \del_{y_1} \mp i \del_{y_2} \right) i \Gamma^{y_1} \Psi_{R\mp}
	- M \Psi_{R\pm} &= 0, 
	\label{eq:6d_Dirac_eq__R} \\
i \Gamma^\mu \del_\mu \Psi_{R\pm} + \left( \del_{y_1} \mp i \del_{y_2} \right) i \Gamma^{y_1} \Psi_{L\mp}
	- M \Psi_{L\pm} &= 0.
	\label{eq:6d_Dirac_eq__L}
}

\subsubsection{KK modes
\label{sec:case-II-KK}}

The mode functions for $\Psi_{R\pm}$ are easily obtained through the BCs at Eq.~{(\ref{eq:BC_case-II})} with suitable normalizations as
\al{
f_{n_1,n_2}(y_1,y_2) = \sqrt{\frac{2}{L_1}} \sqrt{\frac{2}{L_2}}
	\sin\left(\frac{n_1 \pi}{L_1} y_1\right)
	\sin\left(\frac{n_2 \pi}{L_2} y_2\right),
	\label{eq:mode_function_f}
}
with the KK indices $n_1$ and $n_2$.
The functions satisfy the relations
\al{
&f_{n_1, n_2}(y_1,y_2) = 0 \quad \text{at} \ y_1 = 0, L_1 \ {\text{and}}\ y_2 = 0, L_2, 
	\label{eq:BCs_f} \\
&\int_{0}^{L_{1}} \!\!\!\! dy_{1} \int_{0}^{L_{2}} \!\!\!\! dy_{2}
	\left( f_{m_1, m_2}(y_1,y_2) \right)^\ast f_{n_1, n_2}(y_1,y_2) = \delta_{m_1, n_1} \delta_{m_2, n_2}.
	\label{eq:orthonormality_f}
}
The case $n_1$ and/or $n_2$ being zero results in a vanishing profile, which suggests that no zero mode exists as right-handed chiral modes.
Reflecting on this fact, the ranges of $n_1$ and $n_2$ are $n_{1,2} = 1,2,3,\cdots$.
Now, the KK expansion of $\Psi_{R\pm}$ are easily made as
\al{
\Psi_{R\pm}(x,y) = \sum_{n_1 = 1}^\infty \sum_{n_2 = 1}^\infty \psi_{R\pm}^{(n_1, n_2)}(x) f_{n_1, n_2}(y_1,y_2),
	\label{eq:form_of_Psi_Rpm}
}
where ${\psi_{R\pm}^{(n_1, n_2)}(x)}$ are the corresponding 4d chiral fields with right chirality.

The remaining left-handed part {is} described through Eqs.~(\ref{eq:6d_Dirac_eq__R+}) and (\ref{eq:6d_Dirac_eq__R-}) as
\al{
i \Gamma^\mu \del_\mu \Psi_{L\pm} &=
	- \left( i \Gamma^{y_1} \del_{y_1} + i \Gamma^{y_2} \del_{y_2} \right) \Psi_{R\mp} + M \Psi_{R\pm} \notag \\
&=
- i \Gamma^{y_1} \left( \del_{y_1} \mp i \del_{y_2} \right) \Psi_{R\mp} + M \Psi_{R\pm} \notag \\
&=
\sum_{{n_1, n_2} = 1}^{\infty} \left[ (- i \Gamma^{y_1}) \psi_{R\mp}^{(n_1, n_2)}(x) \left( \del_{y_1} \mp i \del_{y_2} \right)
	f_{n_1, n_2}(y) + M \psi_{R\pm}^{(n_1, n_2)}(x) f_{n_1, n_2}(y) \right],
}
which implies the form of $\Psi_{L\pm}$ as
\al{
\Psi_{L\pm}(x,y) &= 
\sum_{{n_1, n_2} = 1}^{\infty} \left[ \eta_{L\pm}^{(n_1, n_2)}(x) \, a_{n_1, n_2} \left( \del_{y_1} \mp i \del_{y_2} \right)
	f_{n_1, n_2}(y) + M \zeta_{L\pm}^{(n_1, n_2)}(x) \, b_{n_1, n_2} f_{n_1, n_2}(y) \right] \notag \\
	&\quad (+ \ \text{zero modes}),
	\label{eq:form_of_Psi_Lpm}
}
where $\eta_{L\pm}^{(n_1, n_2)}$ and $\zeta_{L\pm}^{(n_1, n_2)}$ are 4d left-handed chiral spinors; values of $a_{n_1, n_2}$ and $b_{n_1, n_2}$ should be determined through correct normalizations.
The possible zero-mode part is out of our consideration at this stage, and we come back to the discussion on it later.

Additional information on $a_{n_1, n_2} \, \eta_{L\pm}^{(n_1, n_2)}(x)$ and $b_{n_1, n_2} \, \zeta_{L\pm}^{(n_1, n_2)}(x)$ is extracted when we substitute the form of (\ref{eq:form_of_Psi_Lpm}) in Eqs.~(\ref{eq:6d_Dirac_eq__L+}) and (\ref{eq:6d_Dirac_eq__L-}) at the boundaries, which is
\al{
i \Gamma^{y_1} \zeta_{L\mp}^{(n_1, n_2)}(x) \, b_{n_1, n_2} = M \eta_{L\pm}^{(n_1, n_2)}(x) \, a_{n_1, n_2}.
}
Now, the form of $\Psi_{L\pm}$ can be expressed without $b_{n_1, n_2} \, \zeta_{L\pm}^{(n_1, n_2)}(x)$ as
\al{
\Psi_{L\pm}(x,y) &=
\sum_{{n_1, n_2} = 1}^{\infty} a_{n_1, n_2} \left[ \eta_{L\pm}^{(n_1, n_2)}(x) \left( \del_{y_1} \mp i \del_{y_2} \right)
	f_{n_1, n_2}(y) + M (i \Gamma^{y_1}) \eta_{L{\mp}}^{(n_1, n_2)}(x) f_{n_1, n_2}(y) \right] \notag \\
	&\quad (+ \ \text{zero modes}).
	\label{eq:form_of_Psi_Lpm_2}
}

When we re-substitute (\ref{eq:form_of_Psi_Rpm}) and (\ref{eq:form_of_Psi_Lpm_2}) in Eqs.~(\ref{eq:6d_Dirac_eq__R+})--(\ref{eq:6d_Dirac_eq__L-}), after some manipulations, the relations between $\eta_{L\pm}(x)$ and $\psi_{R\pm}(x)$ are declared as follows,
\al{
a_{n_1, n_2} \, i \Gamma^\mu \del_\mu \eta_{L\mp}^{(n_1, n_2)}(x) + (i \Gamma^{y_1}) \psi_{R\pm}^{(n_1, n_2)}(x) &= 0, 
	\label{eq:case-II_imtermediate_1}\\
i \Gamma^\mu \del_\mu \psi_{R\pm}^{(n_1, n_2)}(x) - a_{n_1, n_2} \, (m_{n_1, n_2})^2 (i \Gamma^{y_1}) \eta_{L\mp}^{(n_1, n_2)}(x) &= 0,
	\label{eq:case-II_imtermediate_2}
}
with
\al{
m_{n_1, n_2} = \sqrt{M^2 + \left( \frac{n_1 \pi}{L_1} \right)^2 + \left( \frac{n_2 \pi}{L_2} \right)^2}.
}
From Eqs.~(\ref{eq:case-II_imtermediate_1}) and (\ref{eq:case-II_imtermediate_2}), with algebraic calculations,
we reach the Klein-Gordon equations,
\al{
[\del^\mu \del_\mu - (m_{n_1, n_2})^2] \psi^{(n_1, n_2)}_{R\pm}(x) &= 0, \notag \\
[\del^\mu \del_\mu - (m_{n_1, n_2})^2] \eta^{(n_1, n_2)}_{L\pm}(x) &= 0,
}
where the form of $a_{n_1, n_2}$ is not yet determined.

By use of the whole information which we have obtained, to {derive} the effective Lagrangian of the KK part is tedious but straightforward.
After a bit lengthy estimation, the following form is derived
\al{
S|_{\text{KK part}} &= \int d^4 x \sum_{{n_1, n_2} = 1}^{\infty}
\Bigg\{
\overline{\psi}^{(n_1, n_2)}_{R+}(x) i \Gamma^\mu \del_\mu \psi^{(n_1, n_2)}_{R+}(x) +
\overline{\psi}^{(n_1, n_2)}_{R-}(x) i \Gamma^\mu \del_\mu \psi^{(n_1, n_2)}_{R-}(x) \notag \\
&\hspace{77pt} +
\overline{\eta}^{(n_1, n_2)}_{L+}(x) i \Gamma^\mu \del_\mu \eta^{(n_1, n_2)}_{L+}(x) +
\overline{\eta}^{(n_1, n_2)}_{L-}(x) i \Gamma^\mu \del_\mu \eta^{(n_1, n_2)}_{L-}(x) \notag \\
&\hspace{77pt} -
m_{n_1, n_2} \overline{\psi}^{(n_1, n_2)}_{R+}(x)(i\Gamma^{y_1}) \eta_{L-}^{(n_1, n_2)}(x) -
m_{n_1, n_2} \overline{\psi}^{(n_1, n_2)}_{R-}(x)(i\Gamma^{y_1}) \eta_{L+}^{(n_1, n_2)}(x) \notag \\
&\hspace{77pt} +
m_{n_1, n_2} \overline{\eta}^{(n_1, n_2)}_{L+}(x)(i\Gamma^{y_1}) \psi_{R-}^{(n_1, n_2)}(x) +
m_{n_1, n_2} \overline{\eta}^{(n_1, n_2)}_{L-}(x)(i\Gamma^{y_1}) \psi_{R+}^{(n_1, n_2)}(x)
\Bigg\},
}
where the factors $a_{n_1, n_2}$ are fixed through the correct normalizations in the $\eta_{L\pm}^{(n_1, n_2)}$'s kinetic terms as
\al{
|a_{n_1, n_2}|^2 = \frac{1}{(m_{n_1, n_2})^2} \quad\to\quad a_{n_1, n_2} = \frac{1}{m_{n_1, n_2}}.
}
Finally, let us convert the above form into the standard shape by the redefinition of the fields of
\al{
\xi_{L\pm}^{(n_1, n_2)}(x) &\equiv (i \Gamma^{y_1}) \eta_{L\mp}^{(n_1, n_2)}(x), \notag \\
\chi_{\pm}^{(n_1, n_2)}(x) &\equiv \psi_{R\pm}^{(n_1, n_2)}(x) + {\xi}_{L\pm}^{(n_1, n_2)}(x),
}
where adopting these bases leads to
\al{
S|_{\text{KK part}}
&= \int d^4 x \sum_{{n_1, n_2} = 1}^{\infty}
\Bigg\{
\overline{\chi}_{+}^{(n_1, n_2)}{(x)} \left[ i \Gamma^\mu \del_\mu - m_{n_1, n_2} \right] \chi_{+}^{(n_1, n_2)}{(x)} +
\overline{\chi}_{-}^{(n_1, n_2)}{(x)} \left[ i \Gamma^\mu \del_\mu - m_{n_1, n_2} \right] \chi_{-}^{(n_1, n_2)}{(x)}
\Bigg\}.
}
Here, in each level of the KK indices, two Dirac fermions ${\chi_{\pm}^{(n_1, n_2)}(x)}$ appear with the corresponding physical mass $m_{n_1, n_2}$.

\subsubsection{zero modes
\label{sec:case-II-zero}}

Here, zero modes {mean massless modes} which should satisfy the following relations
\al{
i \Gamma^\mu \del_\mu \Psi_{L\pm}^{(0)}(x,y) &= 0, 
	\label{eq:zero_mode_condition_Lpm} \\
i \Gamma^\mu \del_\mu \Psi_{R\pm}^{(0)}(x,y) &= 0,
	\label{eq:zero_mode_condition_Rpm}
}
under which the 6d Dirac equations in (\ref{eq:6d_Dirac_eq__R}) and (\ref{eq:6d_Dirac_eq__L}) are reduced to
\al{
\left( \del_{y_1} \mp i \del_{y_2} \right) i \Gamma^{y_1} \Psi_{R\mp}^{(0)}(x,y) - M \Psi_{R\pm}^{(0)}(x,y) &= 0,
	\label{eq:6d_Weyl_eq__R} \\
\left( \del_{y_1} \mp i \del_{y_2} \right) i \Gamma^{y_1} \Psi_{L\mp}^{(0)}(x,y) - M \Psi_{L\pm}^{(0)}(x,y) &= 0,
	\label{eq:6d_Weyl_eq__L}
}
where the superscript ${}^{(0)}$ designates that the fields are zero modes.
Here, right-handed and left-handed {modes} are not entangled in the equations, which is the significant feature
emerging only in the massless mode.

{Under the chiral boundary conditions in Eq.~(\ref{eq:BC_case-II}), no nonvanishing localized profile is possible for the right-handed components within the finite system of the rectangle.}
Thereby, we focus on the left-handed components described by Eq.~(\ref{eq:6d_Weyl_eq__L}).
Different from the right-handed part, no boundary condition is assigned for the left modes.

It is easy to derive the following forms from Eq.~(\ref{eq:6d_Weyl_eq__L}),
\al{
\left[ (\del_{y_1})^2 + (\del_{y_2})^2 - M^2 \right] \Psi_{L\pm}^{(0)}(x,y) = 0.
	\label{eq:6d_KG_eq_zeromode}
}
We note that the massless zero modes of the above equations are suggested to be bound states since the eigenvalues of $(\del_{y_1})^2 + (\del_{y_2})^2$ should be positive, i.e. $M^2$.
In this manuscript, we take the following Ansatz, which may describe the simplest localized solution on the 2d plane,
\al{
\Psi_{L\pm}^{(0)}(x,y) = N_{\pm} \, \xi^{(0)}_{L\pm}(x) e^{a_\pm y_1 + b_\pm y_2},
	\label{eq:Ansatz_for_leftmode}
}
where $a_\pm$ and $b_\pm$ are complex numbers in general which should obey the relation derived from Eq.~(\ref{eq:6d_KG_eq_zeromode}) of
\al{
(a_\pm)^2 + (b_\pm)^2 = M^2{.}
}
{It} is reasonably parametrized as
\al{
a_\pm \equiv M \cos{\theta_\pm},\quad
b_\pm \equiv M \sin{\theta_\pm},\quad \left( \theta_\pm \in \mathbb{C} \right),
}
which is ensured by the trigonometric relation $\cos^2{\theta_{\pm}} + \sin^2{\theta_{\pm}} = 1$ (even though the parameters $\theta_\pm$ are complex).

Substituting the Ansatz shown in Eq.~(\ref{eq:Ansatz_for_leftmode}) in the two equations of (\ref{eq:6d_Weyl_eq__L}) brings us the following relations,
\al{
\theta_+ &= \theta_- \equiv \theta, 
	\label{eq:theta_relation} \\
N_+ \, \xi^{(0)}_{L+}(x) &= N_- \, e^{-i\theta} i \Gamma^{y_1} \xi^{(0)}_{L-}(x), 
	\label{eq:Weyl_relation_1} \\
N_- \, \xi^{(0)}_{L-}(x) &= N_+ \, e^{+i\theta} i \Gamma^{y_1} \xi^{(0)}_{L+}(x),
	\label{eq:Weyl_relation_2}
}
where (\ref{eq:Weyl_relation_1}) or (\ref{eq:Weyl_relation_2}) describes the connection between the `$+$' mode and `$-$' mode.
Now, we find the zero modes $\Psi_{L\pm}^{(0)}(x,y)$ to be of the form\footnote{
We mention that the integrated form over the complex parameter $\theta$
\al{
\Psi_{L+}^{(0)}(x,y) = N \, \xi_{L}^{(0)}(x) \times \int d^2 \theta \, h(\theta) e^{M(\cos{\theta} y_1 + \sin{\theta} y_2)}
}
{with an arbitrary function $h(\theta)$ on $\theta$} is a solution of the equation.
Details of such a generalized case are not touched in this manuscript.
}
\al{
\Psi_{L+}^{(0)}(x,y) &= \sum_{j=1}^{n} N_j \, \xi^{(0)}_{Lj}(x) \, e^{M(\cos{\theta_j} y_1 + \sin{\theta_j} y_2)}, 
	\label{eq:KK_zero_L+} \\
\Psi_{L-}^{(0)}(x,y) &= \sum_{j=1}^{n} N_j \, e^{i\theta_j} (i \Gamma^{y_1}) \, \xi^{(0)}_{Lj}(x) \, e^{M(\cos{\theta_j} y_1 + \sin{\theta_j} y_2)},
	\label{eq:KK_zero_L}
}
where $n$ denotes the number of independent zero modes, which are discriminated by the index $j$.
It is important to note that the number $n$ is not determined at the present stage.

The value of $n$ is fixed through the process of deriving effective action of the zero modes.
After some straightforward calculations, we reach
\al{
S|_{\text{zero-mode part}} &= \int d^4 x \int_{0}^{L_{1}} \!\!\!\! dy_{1} \int_{0}^{L_{2}} \!\!\!\! dy_{2}
\Bigg\{
\overline{\Psi}^{(0)}_{L+}(x,y) i \Gamma^\mu \del_\mu \Psi_{L+}^{(0)}(x,y) +
\overline{\Psi}^{(0)}_{L-}(x,y) i \Gamma^\mu \del_\mu \Psi_{L-}^{(0)}(x,y)
\Bigg\} \notag \\
&= \int d^4 x \
\Bigg\{
\sum_{j=1}^n |N_j|^2 \left( 1 + e^{-i(\theta^\ast_j - \theta_j)} \right) \overline{\xi}^{(0)}_{Lj}(x) i\Gamma^\mu \del_\mu \xi^{(0)}_{Lj}(x) \notag \\
&\hspace{65pt}
\times
\int_{0}^{L_{1}} \!\!\!\! dy_{1} \int_{0}^{L_{2}} \!\!\!\! dy_{2} \,
e^{M\left[ (\cos{\theta_j^\ast} + \cos{\theta_j}) y_1 + (\sin{\theta_j^\ast} + \sin{\theta_j}) y_2 \right]} \notag \\
&\hspace{42pt} +
\sum_{j,k=1, \atop j \not= k}^n N_j^\ast N_k \left( 1 + e^{-i(\theta^\ast_j - \theta_k)} \right) \overline{\xi}^{(0)}_{Lj}(x) i\Gamma^\mu \del_\mu \xi^{(0)}_{Lk}(x) \notag \\
&\hspace{65pt}
\times
\int_{0}^{L_{1}} \!\!\!\! dy_{1} \int_{0}^{L_{2}} \!\!\!\! dy_{2} \,
e^{M\left[ (\cos{\theta_j^\ast} + \cos{\theta_k}) y_1 + (\sin{\theta_j^\ast} + \sin{\theta_k}) y_2 \right]}
\Bigg\}.
}
Here, if the second kind of terms in the above form remains, where kinetic mixing is observed,
the two modes $j$ and $k$ become {dependent}.
Thereby, these terms should vanish, which requests the condition
\al{
1 + e^{-i (\theta^\ast_j - \theta_k)} = 0
}
{that} means in terms of $\{\theta_j\}$
\al{
\theta^\ast_j - \theta_k = \pi \quad (\text{mod} \ 2\pi) \quad \text{for} \ j \not= k.
}
The above formula tells us two important things:
(i) the maximum number of the independent zero modes are {\it two} due to the periodicity.
(ii) the corresponding two angles $\theta_1$ and $\theta_2$ should be correlated as
\al{
\theta_1 = \theta,\quad \theta_2 = \theta^\ast + \pi, \quad (\theta \in \mathbb{C}).
}
Taking account of the issues, we rewrite the form of the KK expansion of the left-handed zero modes in {Eqs.~(\ref{eq:KK_zero_L+}) and (\ref{eq:KK_zero_L})}
\al{
\Psi_{L+}^{(0)}(x,y) &= 
	N_1 \, \xi^{(0)}_{L1}(x) \, e^{M(\cos{\theta} y_1 + \sin{\theta} y_2)} +
	N_2 \, \xi^{(0)}_{L2}(x) \, e^{-M(\cos{\theta^\ast} y_1 + \sin{\theta^\ast} y_2)}, \label{eq:KK_zero_L_1}\\
\Psi_{L-}^{(0)}(x,y) &= 
	N_1 \, e^{i\theta} (i\Gamma^{y_1}) \, \xi^{(0)}_{L1}(x) \, e^{M(\cos{\theta} y_1 + \sin{\theta} y_2)} -
	N_2 \, e^{i\theta^\ast} (i\Gamma^{y_1}) \, \xi^{(0)}_{L2}(x) \, e^{-M(\cos{\theta^\ast} y_1 + \sin{\theta^\ast} y_2)},
	\label{eq:KK_zero_L_2}
}
where the two normalization factors are easily estimated as
\al{
|N_1|^2 &=
	\frac{M^2 (\cos{\theta^\ast} + \cos{\theta}) (\sin{\theta^\ast} + \sin{\theta})}
	{(1 + e^{-i(\theta^\ast - \theta)})(e^{M(\cos{\theta^\ast} + \cos{\theta})L_1} -1)
	 (e^{M(\sin{\theta^\ast} + \sin{\theta})L_2} -1)}, \\
|N_2|^2 &=
	\frac{M^2 (\cos{\theta^\ast} + \cos{\theta}) (\sin{\theta^\ast} + \sin{\theta})}
	{(1 + e^{+i(\theta^\ast - \theta)})(1 - e^{-M(\cos{\theta^\ast} + \cos{\theta})L_1})
	 (1 - e^{-M(\sin{\theta^\ast} + \sin{\theta})L_2})}.
}
{It is noted that $|N_2|^2$ can be obtained by the replacement $\theta \to \theta^\ast + \pi$ in the form of $|N_1|^2$.}
Also, we rerun the form of the zero-mode four-dimensional effective action for convenience
\al{
S|_{\text{zero-mode part}} &= \int d^4 x \
\Bigg\{
\overline{\xi}^{(0)}_{L1}(x) i \Gamma^\mu \del_\mu \xi_{L1}^{(0)}(x) +
\overline{\xi}^{(0)}_{{L2}}(x) i \Gamma^\mu \del_\mu \xi_{L2}^{(0)}(x)
\Bigg\}.
}
We comment on the cross terms between the zero modes and the KK modes appearing in the evaluation of the 4d effective action of $S$ in Eq.~(\ref{eq:6d_action}).
By use of the zero-mode equation in Eqs.~(\ref{eq:6d_Weyl_eq__R}) and (\ref{eq:6d_Weyl_eq__L}) and integration by parts,
we can show that all of such kind of terms vanish and do not contribute.
{Here, let us explicitly {check that no overlap term appears} in the effective action irrespective of the form of the zero modes, which is expected.
We focus on the mixing of the $(n_1, n_2)$-KK state and zero-mode states,
\al{
S|_{\text{mixing}} &\supset \int d^4 x \int_{0}^{L_{1}} \!\!\!\! dy_{1} \int_{0}^{L_{2}} \!\!\!\! dy_{2}
\Bigg\{
\overline{\Psi}^{(n_1, n_2)}_{L\pm}(x,y) i \Gamma^\mu \del_\mu \Psi_{L\pm}^{(0)}(x,y)\notag \\
&\qquad +
\overline{\Psi}^{(n_1, n_2)}_{R\pm}(x,y) \left[ (\partial_{y_1} \mp i \partial_{y_2} ) (i \Gamma^{y_1}) \Psi^{(0)}_{L\mp}(x,y) - M \Psi^{(0)}_{L\pm}(x,y) \right] + \text{h.c.} \Bigg\}.
	\label{eq:KK_zero__mixing_form}
}
The EOMs for wavefunction profile of $\Psi_{L\pm}^{(0)}(x,y)$ in Eq.~(\ref{eq:6d_Weyl_eq__L}) immediately tells us that the second part of (\ref{eq:KK_zero__mixing_form}) vanishes.
Also, after the following deformation with the form in Eq.~{(\ref{eq:form_of_Psi_Lpm_2})},
\al{
\overline{\Psi}^{(n_1, n_2)}_{L\pm}(x,y) i \Gamma^\mu \del_\mu \Psi_{L\pm}^{(0)}(x,y) 
&\sim
\frac{f_{n_1, n_2}^\ast(y)}{m_{n_1, n_2}} \left[ {-} \overline{\eta}^{(n_1,n_2)}_{L\pm}(x) (\partial_{y_1} \pm i \partial_{y_2}) 
- M \overline{\eta}^{(n_1,n_2)}_{L\mp}(x) i \Gamma^{y_1} \right] i \Gamma^\mu \partial_\mu \Psi^{(0)}_{L\pm}(x,y) \notag \\
&=
\frac{f_{n_1, n_2}^\ast(y)}{m_{n_1, n_2}}
\overline{\eta}^{(n_1,n_2)}_{L\pm}(x)
(i \Gamma^\mu \partial_\mu) (i \Gamma^{y_1}) \notag \\
&\qquad \qquad \times 
\left[ {-} (\partial_{y_1} \pm i \partial_{y_2}) i \Gamma^{y_1} \Psi^{(0)}_{L\pm}(x,y)
{+} M \Psi^{(0)}_{L\mp}(x,y) \right] + \text{h.c.}, 
}
where the $\sim$ symbol shows the equivalence up to total derivative terms on $y_{1,2}$,
we recognize that no overlap term emerges from the first part of (\ref{eq:KK_zero__mixing_form}) through Eq.~(\ref{eq:6d_Weyl_eq__L}).
Since the Dirichlet boundary condition is imposed on $f_{n_1, n_2}(y)$, surface terms do not contribute to the effective action.}

Finally, we briefly touch {Case \GkI} defined in Eq.~(\ref{eq:BC_case-I}), where right-handed zero modes can exist.
Since we never use properties coming from eigenvalues of the 4d chirality, except for the chiral boundary conditions which determine which chirality is realized in the zero-mode sector.
Thereby, when we consider {Case \GkI}, exchanging $R$ and $L$ in the results of {Case \GkII} is enough for obtaining corresponding solutions of zero modes and KK modes.

\subsection{Case {\GkIII} --- a Vector-Like Possibility
\label{sec:case-III}}

Different from {Case \GkII}, the present BCs shown in Eq.~(\ref{eq:BC_case-III}) do not distinguish four dimensional chirality, which implies that the lowest energy states are vector-like and massive.
On the other hand, the two-dimensional chirality defined in Eq.~(\ref{eq:eigenstate_Gammay3}) is discriminated, where the `$+$' DOFs are projected out at the boundaries.
Therefore, Dirac equations in the following ``vector-like" forms are suitable for the present analysis, which is given by the summation of Eqs.~(\ref{eq:6d_Dirac_eq__R}) and (\ref{eq:6d_Dirac_eq__L}) as 
\al{
\left[ i \Gamma^\mu \del_\mu -M \right] \Psi_{+}(x,y) + \left( \del_{y_1} - i \del_{y_2} \right) i \Gamma^{y_1} \Psi_{-}(x,y) = 0,
	\label{eq:6d_Dirac_eq_p} \\
\left[ i \Gamma^\mu \del_\mu -M \right] \Psi_{-}(x,y) + \left( \del_{y_1} + i \del_{y_2} \right) i \Gamma^{y_1} \Psi_{+}(x,y) = 0,
	\label{eq:6d_Dirac_eq_m}
}
We note that the following discussions are basically parallel to the previous ones.

\subsubsection{KK modes
\label{sec:case-III-KK}}

Like the previous case, the form of the KK expansion of $\Psi_{+}$ is easily written down,
\al{
\Psi_{+}(x,y) = \sum_{{n_1, n_2} = 1}^\infty \psi_+^{(n_1, n_2)}(x) \, f_{n_1, n_2}(y),
	\label{eq:rectangle_f}
}
where $\psi_+^{(n_1, n_2)}(x)$ is the corresponding 4d fields and the mode functions $f_{n_1, n_2}(y)$ take the same forms as in Eq.~(\ref{eq:mode_function_f}), which fulfill the BCs (\ref{eq:BCs_f}) shown in Eq.~(\ref{eq:BC_case-III}) and the orthonormality (\ref{eq:orthonormality_f}). 
Substituting the above form in Eq.~{(\ref{eq:6d_Dirac_eq_m})} leads to
\al{
\left[ i \Gamma^\mu \del_\mu -M \right] \Psi_{-}(x,y) =
{\sum_{n_1, n_2 = 1}^\infty} (-i\Gamma^{y_1}) \, \psi_{+}^{(n_1, n_2)}(x) (\del_{y_1} +i \del_{y_2}) \, f_{n_1, n_2}(y),
}
which suggests the following form for $\Psi_{-}$
\al{
\Psi_{-}(x,y) &= {\sum_{n_1, n_2 = 1}^\infty} \eta_{-}^{(n_1, n_2)}(x) \, g_{n_1, n_2}(y)
+ (\text{zero modes}), \\
g_{n_1, n_2}(y) &\equiv \frac{1}{\widetilde{m}_{n_1, n_2}} \left( \del_{y_1} + i \del_{y_2} \right) f_{n_1, n_2}(y),
}
where $\widetilde{m}_{n_1, n_2}$ is a normalization constant with mass dimension one.
In this section, we do not touch the zero modes, which is discussed in {subsection~\ref{sec:case-III-zero}}.
The value of $\widetilde{m}_{n_1, n_2}$ is determined through the normalization of the kinetic terms of $\eta_{-}^{(n_1, n_2)}$ via the 6d term $\overline{\Psi}_{-} i \Gamma^\mu \del_\mu \Psi_{-}$ as
\al{
\widetilde{m}_{n_1, n_2} = \sqrt{ \left(\frac{n_1 \pi}{L_1}\right)^2 + \left(\frac{n_2 \pi}{L_2}\right)^2 }.
}
Now, we straightforwardly evaluate the form of the effective action of the KK modes, which is given by
\al{
S|_{\text{KK part}}
&= \int d^4 x \sum_{{n_1, n_2} = 1}^{\infty}
\Bigg\{
\overline{\psi}^{(n_1, n_2)}_{+}(x) (i \Gamma^\mu \del_\mu ) \, \psi_{+}^{(n_1, n_2)}(x) +
\overline{\xi}^{(n_1, n_2)}_{+}(x) (i \Gamma^\mu \del_\mu ) \, \xi_{+}^{(n_1, n_2)}(x) \notag \\
&\hspace{67pt} -
\begin{pmatrix} \overline{\psi}^{(n_1, n_2)}_{+}(x) & \overline{\xi}^{(n_1, n_2)}_{+}(x) \end{pmatrix}
\begin{pmatrix}
M & \widetilde{m}_{n_1, n_2} \\
\widetilde{m}_{n_1, n_2} & -M
\end{pmatrix}
\begin{pmatrix} {\psi}^{(n_1, n_2)}_{+}(x) \\ {\xi}^{(n_1, n_2)}_{+}(x) \end{pmatrix}
\Bigg\},
}
where we refine $\eta^{(n_1, n_2)}_{-}$ as
\al{
\xi_{+}^{(n_1, n_2)} \equiv (i\Gamma^{y_1}) \, \eta_{-}^{(n_1, n_2)}(x).
}
Here, the action is chiral in the sense of the {internal chirality} ($\pm$), while it is vector-like in the 4d chirality point of view.
Similar to {Case \GkII}, two Dirac fermions appear in each level of the KK tower.
The mass eigenvalues of the two types of Dirac states are obtained as
\al{
\pm \sqrt{M^2 + (\widetilde{m}_{n_1, n_2})^2}.
}
After the redefinition of the fields with two-by-two unitary matrices $U_{n_1, n_2}$ such as
\al{
\begin{pmatrix} {\psi'}^{(n_1, n_2)}_{+}(x) \\ {\xi'}^{(n_1, n_2)}_{+}(x) \end{pmatrix}
=
U_{n_1, n_2}
\begin{pmatrix} {\psi}^{(n_1, n_2)}_{+}(x) \\ {\xi}^{(n_1, n_2)}_{+}(x) \end{pmatrix},
}
the mass matrix is diagonalized as
\al{
U_{n_1, n_2}
\begin{pmatrix}
M & \widetilde{m}_{n_1, n_2} \\
\widetilde{m}_{n_1, n_2} & -M
\end{pmatrix}
(U_{n_1, n_2})^\dagger
=
{\begin{pmatrix}
\sqrt{M^2 + (\widetilde{m}_{n_1, n_2})^2} & 0 \\ 
0 & -\sqrt{M^2 + (\widetilde{m}_{n_1, n_2})^2}
\end{pmatrix}}.
}

\subsubsection{zero modes
\label{sec:case-III-zero}}

Here, zero modes mean the physical spectrum obeying the equations
\al{
\left[ i \Gamma^\mu \del_\mu -M \right] \Psi_{+}^{(0)}(x,y) &= 0,
	\label{eq:zero_mode_condition_+} \\
\left[ i \Gamma^\mu \del_\mu -M \right] \Psi_{-}^{(0)}(x,y) &= 0,
	\label{eq:zero_mode_condition_-}
}
where these states are massive states with the common mass eigenvalue $M$ if they exist consistently.
As we pointed out beforehand, we cannot obtain a chiral theory since the present BCs in Eq.~(\ref{eq:BC_case-III}) do not discriminate the 4d chirality.

Under the presence of the above conditions, the 6d Dirac equations take the simplified form
\al{
\left( \del_{y_1} \mp i \del_{y_2} \right) \Psi_{\mp}^{(0)}(x,y) &= 0,
	\label{eq:6d_Weyl_eq__pm}
}
where, different from Eqs.~(\ref{eq:6d_Weyl_eq__R}) and (\ref{eq:6d_Weyl_eq__L}), the massive parameter $M$ does not contribute.
In the language of mode function, (\ref{eq:6d_Weyl_eq__pm}) are {represented} as
\al{
{\left( \del_{y_1} + i \del_{y_2} \right) f_0(y) =  \del_{\bar{z}} f_0(y)} &= 0, 
	\label{eq:holomorphic_eq} \\
{\left( \del_{y_1} - i \del_{y_2} \right) g_0(y) =  \del_{z} g_0(y)} &= 0, 
	\label{eq:anti-holomorphic_eq}
}
where we remind that $f_0$ and $g_0$ are zero-mode eigenfunction of the {internal chirality} being $+$ and $-$, respectively.
{Here, we define {the holomorphic and anti-holomorphic coordinates} and corresponding derivatives as follows,}
\al{
{z \equiv \frac{1}{2}\left( y_1 + i y_2 \right) \,\left( \leftrightarrow \del_{z}       = \del_{y_1} - i \del_{y_2} \right),\quad
\bar{z} \equiv \frac{1}{2}\left( y_1 - i y_2 \right) \,\left( \leftrightarrow \del_{\bar{z}} = \del_{y_1} + i \del_{y_2} \right).}
	\label{eq:holomorphic_coordinate}
}
No zero mode for $f_0$ will exist because $\Psi_{+}$ has to satisfy the Dirichlet BC~(\ref{eq:BC_case-III}) and hence the set $\{ f_{n_1, n_2}(y) ; n_1, n_2 = 1,2,3,\cdots \}$ forms a complete set without zero modes.

On the other hand, $g_0$ could exist and a general solution to (\ref{eq:anti-holomorphic_eq}) would be given by
\al{
{g_{0j}(y) = \phi_j(\overline{z})},
	\label{eq:form_g0j}
}
where $\phi_j(\overline{z})$ are arbitrary anti-holomorphic functions of $\overline{z}$ with the index $j\,(=1,2,\cdots,n)$ specifying independent zero-mode solutions.
For the $n$ number of physical modes are well defined, we may impose the orthonormal condition such as
\al{
\int_{0}^{L_{1}} \!\!\!\! dy_{1} \int_{0}^{L_{2}} \!\!\!\! dy_{2}
\left( \phi_j(\overline{z}) \right)^\ast \phi_k(\overline{z}) = \delta_{jk}.
	\label{eq:conformal_orthonormality}
}
Even though the zero-mode solutions take the generic form, no cross term between zero modes and nonzero KK modes emerges, which is ensured by the equation for $g_{0j}$ in Eq.~(\ref{eq:anti-holomorphic_eq}) with the manipulation of integration by parts over $y_1$ and $y_2${, as discussed concretely in Case \GkII.}

Finally, let us mention the ``opposite" case where $\Psi_{-}(x,y) = 0$ at the boundaries.
Here, no meaningful {zero-mode} solution would exist for $g_0$, while $f_0$ forms arbitrary holomorphic functions of $z$ as follows,
\al{
{f_{0j}(y) = \phi_j({z})}.
	\label{eq:form_f0j}
}

\subsection{Comment on 6d Weyl case
\label{sec:expansion_Weyl}}

{
Here, we briefly comment on the mode functions in the case that a 6d Weyl fermion is considered.
As summarized in Eqs.~(\ref{eq:type-I_condition_6d_Weyl}) and (\ref{eq:type-II_condition_6d_Weyl}), only the type-I and type-II BCs are possible for $\Psi_{\Gamma^7 = +1}$ or $\Psi_{\Gamma^7 = -1}$, which {can be regarded as a 4d Dirac fermion}, {while no other reasonable condition is derived in the type-$\text{\GkIII}$}.
Since the two BCs discriminate 4d chiralities, the zero modes can become chiral.
{In the present Weyl cases}, a nonzero bulk mass parameter $M$ is not allowed for a 6d Weyl fermion.
{In the type-I and type-$\text{\GkII}$ cases, when $M \to 0$, the equations of motion in Eqs.~(\ref{eq:6d_Weyl_eq__R}) and (\ref{eq:6d_Weyl_eq__L}) are reduced to that in Eq.~(\ref{eq:6d_Weyl_eq__pm}) under the constraint $M=0$.
Following the discussion in section~\ref{sec:case-III-zero}, we conclude that the profile of the chiral mode can take arbitrary holomorphic or anti-holomorphic function, depending on the form of corresponding equations.
We mention that the multiplicity of the chiral mode is not determined, as in the case argued in section~\ref{sec:case-III-zero}.}
Both of the choices in the 6d chirality, $+1$ or $-1$, is fine for obtaining a 4d Weyl mode.
Apparently in each level of KK states, a 4d Dirac fermion appears. 
}


\section{{Correspondence to Orbifolds}
\label{sec:orbifold}}

In this section, we argue correspondence between the 6d Dirac theory on a rectangle and that on orbifolds based on the two-dimensional torus $T^2$ {to accomplish a deeper understanding on the theory on a rectangle}.
At first, we glance at the geometry of $T^2$, defined by the two identifications,
\al{
y_1 \sim y_1 + 2L_1,\quad
y_2 \sim y_2 + 2L_2.
}
Here, a choice of the fundamental domain of $T^2$ is
\al{
y_1:\,[-L_1, L_1],\quad
y_2:\,[-L_2, L_2].
}
In the following discussion, we consider the periodic boundary condition for 6d fermions.
The 6d free action of a 6d Dirac fermion on $T^2$ is written down as
\al{
S_{T^2} &= \int \! d^{4} x \int_{-L_1}^{L_{1}} \!\!\!\! dy_{1} \int_{-L_2}^{L_{2}} \!\!\!\! dy_{2} \,
	\overline{\Psi}(x,y) \left( i \Gamma^A \del_A - M \right) \Psi(x,y) 
	\label{eq:6d_action_T2} \\
	&= \int \! d^{4} x \int_{-L_1}^{L_{1}} \!\!\!\! dy_{1} \int_{-L_2}^{L_{2}} \!\!\!\! dy_{2} \,
	\Bigg\{ 
	\overline{\Psi}_{+}(x,y_1,y_2) \, i \Gamma^\mu \del_\mu  \Psi_{+}(x,y_1,y_2) +
	\overline{\Psi}_{-}(x,y_1,y_2) \, i \Gamma^\mu \del_\mu  \Psi_{-}(x,y_1,y_2) \notag \\
	&\phantom{\int \! d^{4} x \int_{-L_1}^{L_{1}} \!\!\!\! dy_{1} \int_{-L_2}^{L_{2}} \!\!\!\! dy_{2} \, \ \ \ }
	+
	\overline{\Psi}_{+}(x,y_1,y_2) \, i \Gamma^{y_1} \del_{z}  \Psi_{-}(x,y_1,y_2) +
	\overline{\Psi}_{-}(x,y_1,y_2) \, i \Gamma^{y_1} \del_{\bar{z}}  \Psi_{+}(x,y_1,y_2) \notag \\
	&\phantom{\int \! d^{4} x \int_{-L_1}^{L_{1}} \!\!\!\! dy_{1} \int_{-L_2}^{L_{2}} \!\!\!\! dy_{2} \, \ \ \ }
	- M
	\left[
	\overline{\Psi}_{+}(x,y_1,y_2) \Psi_{+}(x,y_1,y_2) +
	\overline{\Psi}_{-}(x,y_1,y_2) \Psi_{-}(x,y_1,y_2)
	\right]
	\Bigg\} 
	\label{eq:6d_action_T2_secondline} \\
	&= \int \! d^{4} x \int_{-L_1}^{L_{1}} \!\!\!\! dy_{1} \int_{-L_2}^{L_{2}} \!\!\!\! dy_{2} \,
	\Bigg\{ 
	\overline{\Psi}_{R+}(x,y_1,y_2) \, i \Gamma^\mu \del_\mu  \Psi_{R+}(x,y_1,y_2) +
	\overline{\Psi}_{R-}(x,y_1,y_2) \, i \Gamma^\mu \del_\mu  \Psi_{R-}(x,y_1,y_2) \notag \\
	&\phantom{\int \! d^{4} x \int_{-L_1}^{L_{1}} \!\!\!\! dy_{1} \int_{-L_2}^{L_{2}} \!\!\!\! dy_{2} \, \ \ \ }
	+
	\overline{\Psi}_{L+}(x,y_1,y_2) \, i \Gamma^\mu \del_\mu  \Psi_{L+}(x,y_1,y_2) +
	\overline{\Psi}_{L-}(x,y_1,y_2) \, i \Gamma^\mu \del_\mu  \Psi_{L-}(x,y_1,y_2) \notag \\
	&\phantom{\int \! d^{4} x \int_{-L_1}^{L_{1}} \!\!\!\! dy_{1} \int_{-L_2}^{L_{2}} \!\!\!\! dy_{2} \, \ \ \ }
	+
	\overline{\Psi}_{R+}(x,y_1,y_2) \, i \Gamma^{y_1} \del_{z}  \Psi_{L-}(x,y_1,y_2) +
	\overline{\Psi}_{L-}(x,y_1,y_2) \, i \Gamma^{y_1} \del_{\bar{z}}  \Psi_{R+}(x,y_1,y_2) \notag \\
	&\phantom{\int \! d^{4} x \int_{-L_1}^{L_{1}} \!\!\!\! dy_{1} \int_{-L_2}^{L_{2}} \!\!\!\! dy_{2} \, \ \ \ }
	+
	\overline{\Psi}_{L+}(x,y_1,y_2) \, i \Gamma^{y_1} \del_{z}  \Psi_{R-}(x,y_1,y_2) +
	\overline{\Psi}_{R-}(x,y_1,y_2) \, i \Gamma^{y_1} \del_{\bar{z}}  \Psi_{L+}(x,y_1,y_2) \notag \\
	&\phantom{\int \! d^{4} x \int_{-L_1}^{L_{1}} \!\!\!\! dy_{1} \int_{-L_2}^{L_{2}} \!\!\!\! dy_{2} \, \ \ \ }
	- M
	\Big[
	\overline{\Psi}_{R+}(x,y_1,y_2) \Psi_{L+}(x,y_1,y_2) +
	\overline{\Psi}_{L+}(x,y_1,y_2) \Psi_{R+}(x,y_1,y_2) \notag \\
	&\phantom{\int \! d^{4} x \int_{-L_1}^{L_{1}} \!\!\!\! dy_{1} \int_{-L_2}^{L_{2}} \!\!\!\! dy_{2} \,
	          \qquad \ \ } +
	\overline{\Psi}_{R-}(x,y_1,y_2) \Psi_{L-}(x,y_1,y_2) +
	\overline{\Psi}_{L-}(x,y_1,y_2) \Psi_{R-}(x,y_1,y_2)
	\Big]
	\Bigg\},
	\label{eq:6d_action_T2_thirdline}
}
where we {used} the complex coordinate defined in Eq.~(\ref{eq:holomorphic_coordinate}) and $(i \Gamma^{y_1})^2 = I_8$.
{Here, we decomposed $\Psi$ into the eigenstates of $\mathcal{P}_{R/L}$ and $\mathcal{P}_{\pm}$.}
The mode functions on $T^2$ (without Scherk--Schwarz twist) take the generic form
\al{
\exp\left( \frac{i \pi n_1}{L_1} y_1 \right) \exp\left( \frac{i \pi n_2}{L_2} y_2 \right),
	\label{eq:T2_modefunction}
}
where $n_1$ and $n_2$ ($= 0,\,\pm1,\,\pm2,\cdots$) are KK indices and we do not take care of correct normalization of wavefunctions {throughout this section}.

\subsection{$T^2/Z_N$ Twisted Orbifold
\label{sec:twisted}}

In this part, we address a direction of the twisted orbifolds on $T^2$, namely $T^2/Z_2$, $T^2/Z_3$, $T^2/Z_4$ and $T^2/Z_6$.
The $Z_N$ ($N=2,3,4,6$) operations are defined as the identifications of the points on $T^2$ under the rotation on the $y_1y_2$ plane,\footnote{
For $T^2/Z_3$, $T^2/Z_4$ and $T^2/Z_6$, {the condition $L_1 = L_2$ is required to keep the rotations well defined}.
}
\al{
\begin{pmatrix} y'_{1} \\ y'_{2} \end{pmatrix}
&=
\begin{pmatrix}
\cos{\theta} & \sin{\theta} \\
-\sin{\theta} & \cos{\theta}
\end{pmatrix}
\begin{pmatrix} y_{1} \\ y_{2} \end{pmatrix}\quad
(\theta = 2\pi/N),
}
where subsequently the 6d spinor fields are also rotated as designated by the matrix $R_y$ following the corresponding generator $L_y$ for 6d spinors,
\al{
R_y &= e^{-i \theta L_y}
	 = \cos\left(\frac{\theta}{2}\right) I_8 -i \sin\left(\frac{\theta}{2}\right) \Gamma^y , \quad L_y \equiv \frac{i}{4} \left[ \Gamma^{y_1}, \Gamma^{y_2} \right].
	 \label{eq:6d_spinor_rotation_generator}
}
The commutativity $[R_y, \Gamma^7] = 0$ tells us that the following {$Z_N$ parity assignments} are possible,
\al{
\Psi_{\Gamma^7=\pm1}(x, y'_1, y'_2) =
	\begin{cases}
	\eta_{Z_2}^{(\pm)} \begin{pmatrix} I_4 & 0 \\ 0 & -I_4 \end{pmatrix} \Psi_{\Gamma^7=\pm1}(x, y_1, y_2) &
		\text{in } T^2/Z_2, \\
	\eta_{Z_3}^{(\pm)} \begin{pmatrix} I_4 & 0 \\ 0 & e^{i 2\pi/3} I_4 \end{pmatrix} \Psi_{\Gamma^7=\pm1}(x, y_1, y_2) &
		\text{in } T^2/Z_3, \\
	\eta_{Z_4}^{(\pm)} \begin{pmatrix} I_4 & 0 \\ 0 & e^{i \pi/2} I_4 \end{pmatrix} \Psi_{\Gamma^7=\pm1}(x, y_1, y_2) &
		\text{in } T^2/Z_4, \\
	\eta_{Z_6}^{(\pm)} \begin{pmatrix} I_4 & 0 \\ 0 & e^{i \pi/3} I_4 \end{pmatrix} \Psi_{\Gamma^7=\pm1}(x, y_1, y_2) &
		\text{in } T^2/Z_6,
	\end{cases}
}
with intrinsic $Z_{N}$ parities for $\Psi_{\Gamma^7=\pm1}$, $\eta_{Z_{N}}^{(\pm)}$ which take one of the values of the $N$-th roots of unity, $(e^{2 \pi i/N})^j$ ($j = 0,\cdots,N-1$).

A point is that we cannot obtain chiral zero mode from a 6d Dirac fermion on $T^2/Z_N$ if $\eta^{(+)}_{Z_N} = \eta^{(-)}_{Z_N}$, where the lowest mode is a 4d Dirac state in the cases of $\eta_{Z_{N}}^{(\pm)} = 1$ or $e^{-2 \pi i/N}$.
Two chiral modes appear if different BCs are imposed for $\Psi_{\Gamma^7=\pm1}$, namely {$\{\eta_{Z_{N}}^{(+)}, \eta_{Z_{N}}^{(-)}\} = \{1,e^{-2 \pi i/N}\}$} (for right modes) or {$\{\eta_{Z_{N}}^{(+)}, \eta_{Z_{N}}^{(-)}\} = \{e^{-2 \pi i/N},1\}$} (for left modes), where the {zero mode} spectrum is {the same as} that of Type I and Type \GkII, respectively.
On the other hand, a notable difference is also found on the 6d bulk mass term.
When $\eta^{(+)}_{Z_N} \not= \eta^{(-)}_{Z_N}$, the term is forbidden by the $Z_N$ symmetry.\footnote{
{We could add a kink-like mass term that is consistent with the $Z_{N}$ parity as introduced in the $S^1/Z_2$ geometry (e.g., in {Ref.}~\cite{Barbieri:2002ic}).}
}
This fact means that the lowest modes cannot take localized profiles like in Eqs.~(\ref{eq:KK_zero_L+}) and (\ref{eq:KK_zero_L}), which should be constant.
The KK mode functions also take different shapes from those on a rectangle, e.g. in $T^2/Z_2$,
\al{
\cos\left( \frac{\pi n_1 y_1}{L_1} + \frac{\pi n_2 y_2}{L_2} \right)\ \text{for $Z_2$ even},\quad
\sin\left( \frac{\pi n_1 y_2}{L_1} + \frac{\pi n_2 y_2}{L_2} \right)\ \text{for $Z_2$ odd},
} 
where one refers to Eq.~(\ref{eq:mode_function_f}).
{This fact implies that if we introduce interaction terms with other 6d fields, then magnitudes of 4d coupling constants of interaction terms in the $T^2/Z_N$ twisted orbifold model will be different from those of our model.
Thus, the $T^2/Z_N$ twisted orbifold models turn out not to realize the 6d Dirac theory on a rectangle.
This conclusion also can be seen from the fact that the $T^2/Z_N$ twisted orbifolds are geometrically different from a rectangle.}

\subsection{$T^2/(Z_2 \times Z'_2)$ Reflectional Orbifold 
\label{sec:Z2Z2}}

Next, we argue the possibility of the $T^2/(Z_2 \times T'_2)$ reflectional orbifold, where the following reflections are imposed,
\al{
 Z_2:& \ (y_1, y_2) \rightarrow (-y_1, y_2) \quad\longleftrightarrow\quad (z, \bar{z}) \rightarrow {(-\bar{z}, -z)}, \\
Z'_2:& \ (y_1, y_2) \rightarrow (y_1, -y_2) \quad\longleftrightarrow\quad (z, \bar{z}) \rightarrow {(\bar{z}, z)}.
}
In the present setup, the fundamental domain of $(y_1, y_2)$, which is shrunk by the projections, can be chosen as $y_1$:\,$[0,L_1]$ and $y_2$:\,$[0,L_2]$, which corresponds to the rectangle one.
In such orbifold constructions with two different identifications, consistent conditions for 6d fermions may take rather nontrivial forms.

\subsubsection{A Simple Trial, {failed}
\label{sec:foiled_case}}

The first expression of Eq.~(\ref{eq:6d_action_T2}) tells us the conditions on transformations of fermion requested by the $Z_2$ symmetries.
When a 6d Dirac fermion $\Psi$ is transformed as
\al{
 Z_2:& \ \Psi(x,-y_1,y_2) = G_1 \Psi(x,y_1,y_2) \quad \text{with} \ (G_1)^2 = I_8, \\
Z'_2:& \ \Psi(x,y_1,-y_2) = G_2 \Psi(x,y_1,y_2) \quad \text{with} \ (G_2)^2 = I_8,
} 
all of the following relations should be realized to keep the original action to be {invariant},
\al{
Z_2:
\begin{cases}
\Gamma^0 G_1^\dagger \Gamma^0 \Gamma^\mu G_1   &= +\Gamma^\mu, \\
\Gamma^0 G_1^\dagger \Gamma^0 \Gamma^{y_1} G_1 &= -\Gamma^{y_1}, \\
\Gamma^0 G_1^\dagger \Gamma^0 \Gamma^{y_2} G_1 &= +\Gamma^{y_2}, \\
\Gamma^0 G_1^\dagger \Gamma^0 I_8 G_1 \times M &= +I_8 \times M,
\end{cases}
\qquad
Z'_2:
\begin{cases}
\Gamma^0 G_2^\dagger \Gamma^0 \Gamma^\mu G_2   &= +\Gamma^\mu, \\
\Gamma^0 G_2^\dagger \Gamma^0 \Gamma^{y_1} G_2 &= +\Gamma^{y_1}, \\
\Gamma^0 G_2^\dagger \Gamma^0 \Gamma^{y_2} G_2 &= -\Gamma^{y_2}, \\
\Gamma^0 G_2^\dagger \Gamma^0 I_8 G_2 \times M &= +I_8 \times M.
	\label{eq:simple_Z2Z2}
\end{cases} 
}
The choice of $G_1$ and $G_2$,
\al{
G_1 = i\,\Gamma^{y_1},\quad
G_2 = i\,\Gamma^{y_2},
}
fulfills the requirements in Eq.~(\ref{eq:simple_Z2Z2}) when $M=0$.
However, the present case cannot be defined well since the two operations are not commutative, which is recognized by
\al{
{[G_1, G_2] = 2 G_1 G_2 \not= 0} \quad \left( \text{since } \{G_1, G_2\} = 0 \right).
}
Then, we should abandon this possibility.

\subsubsection{Consistent Configuration, corresponding to Case $\text{\GkII}$
\label{sec:Z2Z2_caseII}}

Here, we explore a consistent configuration where two left-handed zero modes emerge, which is derived in Case \GkII.
A key point is to focus on the last form of the 6d action in Eq.~(\ref{eq:6d_action_T2}). 
The bilinear terms that contains the matrix $\Gamma^{y_1}$ is invariant when the following conditions are considered,
\al{
Z_2 (z \leftrightarrow -\bar{z}):&
\begin{cases}
\Psi_{R\pm}(x,-y_1,y_2) &= - \Psi_{R\pm}(x,y_1,y_2), \\
\del_{z} \Psi_{L+}(x,-y_1,y_2) &= + \del_{\bar{z}} \Psi_{L+}(x,y_1,y_2), \\
\del_{\bar{z}} \Psi_{L-}(x,-y_1,y_2) &= + \del_{z} \Psi_{L-}(x,y_1,y_2),
\end{cases} 
	\label{eq:Z2Z2_caseII_condition_1} \\
Z'_2 (z \leftrightarrow \bar{z}):&
\begin{cases}
\Psi_{R\pm}(x,y_1,-y_2) &= - \Psi_{R\pm}(x,y_1,y_2), \\
\del_{z} \Psi_{L+}(x,y_1,-y_2) &= - \del_{\bar{z}} \Psi_{L+}(x,y_1,y_2), \\
\del_{\bar{z}} \Psi_{L-}(x,y_1,-y_2) &= - \del_{z} \Psi_{L-}(x,y_1,y_2),
\end{cases} 
	\label{eq:Z2Z2_caseII_condition_2}
}
where the factor $(-1)$ would appear even times in every term of the last form {(\ref{eq:6d_action_T2_thirdline})} of Eq.~(\ref{eq:6d_action_T2}), {irrespective of the part of the 6d Dirac mass term}.
We mention that these conditions do not contain $\Gamma^{y_1}$ and $\Gamma^{y_2}$, and then no unwanted minus sign from exchanging gamma matrices would emerge.
Thus, it is apparent that the two operations are commutative.
We note that the $Z_2$ conditions are rewritten as follows,
\al{
Z_2 (z \leftrightarrow -\bar{z}):&
\begin{cases}
\mathcal{P}_{R} \Psi(x,-y_1,y_2) &= - \mathcal{P}_{R} {\Psi} (x,y_1,y_2), \\
\mathcal{P}_{L} \left( I_8 \del_{y_1} -i \Gamma^y \del_{y_2} \right) \Psi(x,-y_1,y_2) &= + 
\mathcal{P}_{L} \left( I_8 \del_{y_1} +i \Gamma^y \del_{y_2} \right)\Psi(x,y_1,y_2),
\end{cases} \\
Z'_2 (z \leftrightarrow \bar{z}):&
\begin{cases}
\mathcal{P}_{R} \Psi(x,y_1,-y_2) &= - \mathcal{P}_{R} {\Psi} (x,y_1,y_2), \\
\mathcal{P}_{L} \left( I_8 \del_{y_1} -i \Gamma^y \del_{y_2} \right) \Psi(x,y_1,-y_2) &= - 
\mathcal{P}_{L} \left( I_8 \del_{y_1} +i \Gamma^y \del_{y_2} \right)\Psi(x,y_1,y_2).
\end{cases} 
}
It is proved that, except for the {6d} Dirac mass terms, all of the terms of $S_{T^2}$ in Eq.~(\ref{eq:6d_action_T2}) is invariant under the $Z_2 \times Z'_2$ orbifolding.
We provide the proof of the invariance of the action in Eq.~(\ref{eq:6d_action_T2}) under the $Z_2 \times Z'_2$ operation in appendix~\ref{appendix:Z2Z2}.

To know parities of the fermion profiles under the reflections $y_1 \to -y_1$ and $y_2 \to -y_2$, it is very convenient to express the two $Z_2$ conditions in the following way,
\al{
Z_2 (y_1 \to -y_1):&
\begin{cases}
\Psi_{R\pm}(x,y_1,y_2)|_{y_1 \to -y_1} &= - \Psi_{R\pm}(x,y_1,y_2), \\
\del_{\bar{z}} \Psi_{L+}(x,y_1,y_2)|_{y_1 \to -y_1} &= - \del_{\bar{z}} \Psi_{L+}(x,y_1,y_2), \\
\del_{z} \Psi_{L-}(x,y_1,y_2)|_{y_1 \to -y_1}       &= - \del_{z} \Psi_{L-}(x,y_1,y_2),
\end{cases} \\
Z'_2 (y_2 \to -y_2):&
\begin{cases}
\Psi_{R\pm}(x,y_1,y_2)|_{y_2 \to -y_2} &= - \Psi_{R\pm}(x,y_1,y_2) (x,y_1,y_2), \\
\del_{\bar{z}} \Psi_{L+}(x,y_1,y_2)|_{y_2 \to -y_2} &= - \del_{\bar{z}} \Psi_{L+}(x,y_1,y_2), \\
\del_{z} \Psi_{L-}(x,y_1,y_2)|_{y_2 \to -y_2}       &= - \del_{z} \Psi_{L-}(x,y_1,y_2).
\end{cases}
}
At first, we easily recognize that the profiles of $\Psi_{R\pm}$, $\del_{\bar{z}} \Psi_{L+}$ and $\del_{z} \Psi_{L-}$ are odd under the two reflections, $y_1 \to -y_1$ and $y_2 \to -y_2$, and thereby their values become zero at $(y_1, y_2) = (0,0)$.
{Here, we advert to the fact that the possibility of the geometry $T^2/(Z_2 \times Z'_2)$ was pointed out as $T^2/D_2$ in the work for classifying $S^1$-based (in 5d) and $T^2$-based (in 6d) orbifolds in Ref.~\cite{Nilse:2006jv}.
On the other hand, to the best of our knowledge, the way of a realization of the $Z_2 \times Z'_2$ orbifold condition by use of derivatives for 6d (Dirac) fermions is proposed for the first time on this manuscript.}

Combined with the (assumed) periodicity of mode functions, we reach the conditions at the circumference of the fundamental region of $T^2/(Z_2 \times Z'_2)$,
\al{
\Psi_{R\pm}(x,y_1,y_2)              &= 0  & \text{at } y_1 = 0,\,L_1 \text{ and } y_2 = 0,\,L_2, \\
\del_{\bar{z}} \Psi_{L+}(x,y_1,y_2) &= 0  & \text{at } y_1 = 0,\,L_1 \text{ and } y_2 = 0,\,L_2, \\
\del_{z} \Psi_{L-}(x,y_1,y_2)       &= 0  & \text{at } y_1 = 0,\,L_1 \text{ and } y_2 = 0,\,L_2,
}
which corresponds to Case $\text{\GkII}$ on a rectangle (Type-\GkII-$y_1$ BC in Eq.~(\ref{eq:type-II_condition_6d_y1}) and Type-\GkII-$y_2$ BC in Eq.~(\ref{eq:type-II_condition_6d_y2}), respectively).

We comment on mode functions of $T^2/(Z_2 \times Z'_2)$.
The form on $T^2$ in Eq.~(\ref{eq:T2_modefunction}) and the above boundary conditions immediately lead to
\al{
\text{for } \Psi_{R\pm},&\quad \sin\left( \frac{\pi n_1}{L_1} y_1 \right) \sin\left( \frac{\pi n_2}{L_2} y_2 \right), 
	\label{eq:Rpm_mode_T2Z2Z2} \\
\text{for } \Psi_{L\pm},&\quad 
	\begin{cases}
	\frac{\pi n_1}{L_1} \cos\left( \frac{\pi n_1}{L_1} y_1 \right) \sin\left( \frac{\pi n_2}{L_2} y_2 \right) \mp i \
	\frac{\pi n_2}{L_2} \sin\left( \frac{\pi n_1}{L_1} y_1 \right) \cos\left( \frac{\pi n_2}{L_2} y_2 \right)
		& \text{for } (n_1,n_2) \not= (0,0), \\
	\text{constant}
		& \text{for } (n_1,n_2) = (0,0),
	\label{eq:Lpm_mode_T2Z2Z2}
	\end{cases}
}
where the forms of the KK mode functions are completely the same with those on the rectangle.
We note that {the independent range of $(n_1, n_2)$ is shrunk as $n_1,\,n_2 = 0,\,+1,\,+2,\,\cdots$ from $n_1,\,n_2 = 0,\,\pm1,\,+\pm2,\,\cdots$ from that in $T^2$}, where the two modes $(n_1, n_2) =$ $(1,0)$ and $(0,1)$ are absent since the mode functions vanish.
Here, the existence of two left-handed zero modes is explicitly shown, but being different from the rectangle case, the profile should be constant.
This is because the {6d} Dirac mass term should vanish when we impose the $Z_2 \times Z'_2$ condition and therefore a finite $M$ cannot contribute to wavefunctions.

Finally, we briefly mention the correspondence to Case I on a rectangle, where two right-handed chiral zero modes come out.
Because the 4d chirality and the internal chirality are determined independently, the simple exchange of $R \leftrightarrow L$ is enough to obtain the corresponding situation on $T^2/(Z_2 \times Z'_2)$ from the discussion developed in this section.

\subsubsection{Consistent Configuration, corresponding to Case $\text{\GkIII}$
\label{sec:Z2Z2_caseIII}}

Next, we consider the $T^2/(Z_2 \times Z'_2)$ orbifold corresponding Case $\text{\GkIII}$, where the internal chirality is discriminated by the boundary of a rectangle.
Referring to the second form~{(\ref{eq:6d_action_T2_secondline})} of $S_{T^2}$ in Eq.~(\ref{eq:6d_action_T2}) and the way of constructing the $Z_2 \times Z'_2$ condition in the {previous} chiral case straightaway leads to the conditions,
\al{
Z_2 (z \leftrightarrow -\bar{z}):&
\begin{cases}
\Psi_{+}(x,-y_1,y_2) &= - \Psi_{+}(x,y_1,y_2), \\
\del_{\bar{z}} \Psi_{-}(x,-y_1,y_2) &= + \del_{z} \Psi_{-}(x,y_1,y_2),
\end{cases} 
	\label{eq:Z2Z2_caseIII_condition_1} \\
Z'_2 (z \leftrightarrow \bar{z}):&
\begin{cases}
\Psi_{+}(x,y_1,-y_2) &= - \Psi_{+}(x,y_1,y_2), \\
\del_{\bar{z}} \Psi_{-}(x,y_1,-y_2) &= - \del_{z} \Psi_{-}(x,y_1,y_2),
	\label{eq:Z2Z2_caseIII_condition_2}
\end{cases} 
}
or in the 6d-manifest form
\al{
Z_2 (z \leftrightarrow -\bar{z}):&
\begin{cases}
\mathcal{P}_{+} \Psi(x,-y_1,y_2) &= - \mathcal{P}_{+} \Psi(x,y_1,y_2), \\
\mathcal{P}_{-} \del_{\bar{z}} \Psi(x,-y_1,y_2) &= + \mathcal{P}_{-} \del_{z} \Psi(x,y_1,y_2),
\end{cases} \\
Z'_2 (z \leftrightarrow \bar{z}):&
\begin{cases}
\mathcal{P}_{+} \Psi(x,y_1,-y_2) &= - \mathcal{P}_{+} \Psi(x,y_1,y_2), \\
\mathcal{P}_{-} \del_{\bar{z}} \Psi(x,y_1,-y_2) &= - \mathcal{P}_{-} \del_{z} \Psi(x,y_1,y_2).
\end{cases} 
}
We can easily check that under the transformation, every term of $S_{T^2}$ is invariant, {\it including the 6d bulk mass term}.
Different from the {previous} chiral case, a nonzero $M$ is still consistent with the imposed discrete symmetry, like Case $\text{\GkIII}$ on a rectangle (ref.~appendix~\ref{appendix:Z2Z2}).

Also like as the previous discussion, the reworded conditions
\al{
Z_2 (y_1 \to -y_1):&
\begin{cases}
\Psi_{+}(x,y_1,y_2)|_{y_1 \to -y_1} &= - \Psi_{+}(x,y_1,y_2), \\
\del_{z} \Psi_{-}(x,y_1,y_2)|_{y_1 \to -y_1} &= - \del_{z} \Psi_{-}(x,y_1,y_2),
\end{cases} \\
Z'_2 (y_2 \to -y_2):&
\begin{cases}
\Psi_{{+}}(x,y_1,y_2)|_{y_2 \to -y_2} &= - \Psi_{{+}}(x,y_1,y_2) (x,y_1,y_2), \\
\del_{z} \Psi_{-}(x,y_1,y_2)|_{y_2 \to -y_2} &= - \del_{z} \Psi_{-}(x,y_1,y_2),
\end{cases}
}
immediately tells us the BCs at the circumference of the fundamental region of $T^2/(Z_2 \times Z'_2)$
\al{
\Psi_{+}(x,y_1,y_2)                 &= 0  & \text{at } y_1 = 0,\,L_1 \text{ and } y_2 = 0,\,L_2, \\
\del_{z} \Psi_{-}(x,y_1,y_2)        &= 0  & \text{at } y_1 = 0,\,L_1 \text{ and } y_2 = 0,\,L_2.
}

Now, the forms of mode functions of $\Psi_{+}$ and $\Psi_{-}$ correspond to that in Eq.~(\ref{eq:Rpm_mode_T2Z2Z2}) [for $\Psi_{R\pm}$ in the previous case] and one in Eq.~(\ref{eq:Lpm_mode_T2Z2Z2}) [for $\Psi_{L-}$ in the previous case], respectively.
Also in the present case with nonzero bulk mass, only the constant profile is possible {in the lowest mass states}.
Like in the previous discussion, we find a significant difference on the profile of the lowest mode (with a nonzero mass eigenvalue:\,$M$).


\section{Miscellaneous Issues
\label{sec:misc}}

In this section, we provide several comments on the configurations under the BCs of {Case \GkII} [in Eq.~(\ref{eq:BC_case-II})] and {Case \GkIII} [in Eq.~(\ref{eq:BC_case-III})] obtained in the previous section.
At first, let us summarize the mass spectrum, where concrete information is found in Tables~\ref{tab:case-II} (for {Case \GkII}) and \ref{tab:case-III} (for {Case \GkIII}).

\renewcommand{\arraystretch}{2.2}
\begin{table}[tbh]
\centering
\begin{tabular}{c|c|c|c} \hline
type & fields & Dirac/Weyl & {masses} \\ \hline \hline
\multirow{2}{*}{KK modes} & ${\chi}^{(n_1, n_2)}_{+}(x)$ & Dirac &  $\sqrt{M^2 + \left( \frac{n_1 \pi}{L_1} \right)^2 + \left( \frac{n_2 \pi}{L_2} \right)^2}$ \ $(n_{1}, n_{2} = 1,2,\cdots)$ \\ \cline{2-4}
& ${\chi}^{(n_1, n_2)}_{-}(x)$ & Dirac &  $\sqrt{M^2 + \left( \frac{n_1 \pi}{L_1} \right)^2 + \left( \frac{n_2 \pi}{L_2} \right)^2}$ \ $(n_{1}, n_{2} = 1,2,\cdots)$ \\ \hline\hline
zero modes & ${\xi}_{L1}^{(0)}(x)$, ${\xi}_{L2}^{(0)}(x)$ & left-Weyl & $0$ \\ \hline
\end{tabular}
\caption{Summary of the 4d mass spectrum via the 6d Dirac fermion $\Psi$ under BCs of {Case \GkII} in Eq.~(\ref{eq:BC_case-II}).}
\label{tab:case-II}
\end{table}
\renewcommand{\arraystretch}{1.0}

\renewcommand{\arraystretch}{2.2}
\begin{table}[tbh]
\centering
\begin{tabular}{c|c|c|c} \hline
type & fields & Dirac/Weyl & {masses} \\ \hline \hline
\multirow{2}{*}{KK modes} & ${\psi'}^{(n_1, n_2)}_{+}(x)$ & Dirac &  $\sqrt{M^2 + \left( \frac{n_1 \pi}{L_1} \right)^2 + \left( \frac{n_2 \pi}{L_2} \right)^2}$ \ $(n_{1}, n_{2} = 1,2,\cdots)$ \\ \cline{2-4}
& ${\xi'}^{(n_1, n_2)}_{+}(x)$ & Dirac &  $\sqrt{M^2 + \left( \frac{n_1 \pi}{L_1} \right)^2 + \left( \frac{n_2 \pi}{L_2} \right)^2}$ \ $(n_{1}, n_{2} = 1,2,\cdots)$ \\ \hline\hline
zero modes & ${\eta}^{(0)}_{+j}(x)$ & Dirac &  $M$ $(j=1,2,\cdots,n)$ \\ \hline
\end{tabular}
\caption{Summary of the 4d mass spectrum via the 6d Dirac fermion $\Psi$ under BCs of {Case \GkIII} in Eq.~(\ref{eq:BC_case-III}).}
\label{tab:case-III}
\end{table}
\renewcommand{\arraystretch}{1.0}
\noindent
The spectrum of the KK modes takes the same form, where two Dirac particles appear in each pair of the KK indices $n_1$ and $n_2$ with the common mass.

On the other hand, the structure of the zero modes is completely different.
When we take the boundary conditions which discriminate 4d chirality as {Case \GkII}, chiral fermions are realized as the lowest energy states as in the similar situation in 5d, where the condition $\Psi_{R}^{\text{(5d)}}(x,y) = 0$ is imposed for a 5d Dirac fermion $\Psi_{R}^{\text{(5d)}}$ at the boundaries of an interval.
An important difference between the 5d (on an interval) and the 6d (on a rectangle) is found at the number of the realized chiral zero modes, where {one is in the 5d and two is in the 6d}.
A simple way to understand the difference is that a 6d Dirac fermion contains the twice DOFs compared with that in 5d.
Under the specific Ansatz in Eq.~(\ref{eq:Ansatz_for_leftmode}), we reconfirmed the above simple understanding by discussing how many zero modes can be independent each other, where the answer which we obtained is also two.

Another fascinating aspect is found in the specific solution via the Ansatz in Eq.~(\ref{eq:Ansatz_for_leftmode}), where a complex angle parameter is not determined through the BCs in Eq.~(\ref{eq:BC_case-II}) since the form of the left-handed zero modes are not restricted by them.
After solutions become free from the information on the boundaries, the rectangle looks the two-dimensional Euclidian space for the solutions, and the symmetry under the (complexified) two-dimensional rotation is spontaneously realized inside the form of the solutions as in {Eqs.~(\ref{eq:Weyl_relation_1}) and (\ref{eq:Weyl_relation_2})}.
Note that in the 6d action~(\ref{eq:6d_action}), the existence of the boundaries is manifest and this rotational symmetry is explicitly broken.
In other words, when the value of the angle $\theta$ is different, theories become different.
It is noted that this symmetry can be addressed in a generic manner by focusing on the covariance of the Dirac equation for the zero modes.
A discussion for this subject is ready in appendix~\ref{appendix:spinor_rotation}.
We mention that the value of $\theta$ does not deform mass spectrum, while it affects overlap integrals among a pair of the fermion and other bulk fields with localized profiles.
In this sense, we can conclude that the value of $\theta$ is physical.
{See appendix~\ref{sec:theta} for a concrete discussion on a possible situation that the value of $\theta$ becomes physical.}
For better understanding, we also comment on the situation on an interval.
Here, left-handed modes also cannot feel the presence of the boundaries (when we impose the condition $\Psi_{R}^{\text{(5d)}}(x,y) = 0$ at the two boundaries), but the co-dimension of the extra space is just one and no such enhancement of a rotational symmetry occurs.\footnote{
The general issue that an energy eigenvalue cannot be degenerated in boundary-less one-dimensional quantum mechanical systems, also ensure that one independent mode appears in the 5d case.
Situations are changed if multiple boundary points exist in 5d, where degenerated zero modes become possible (see e.g.~{\cite{Fujimoto:2012wv,Fujimoto:2013ki,Fujimoto:2014fka}}).
}
{As a generalization, when we introduce $n$ 6d Dirac fermions with suitable BCs, {$2n$} numbers of chiral fermions are obtained.
If these right-handed and left-handed chiral fermions couple to a scalar whose vacuum expectation value is $y_1$- and/or $y_2$-position dependent, a part of exotic particles can be very heavy, keeping three particles to be still around GeV scale.}

In {Case \GkIII}, we saw a more drastic situation.
When the zero-mode equations are free from the information on the boundaries and the dimensionful parameter $M$ does not appear in the equations, the 2d rotational symmetry observed at the zero modes is eventually promoted to the 2d conformal symmetry, where the (anti-)holomorphy restricts the form of the zero mode functions as in Eq.~(\ref{eq:form_g0j}) or (\ref{eq:form_f0j}).
Here, the number of such zero modes is not fixed within the discussion done in the manuscript.
At least, the orthonormality in Eq.~(\ref{eq:conformal_orthonormality}) would be imposed for defining the conformal zero modes as physically independent states.
We point out that when we take the limit $M \to 0$ in {Case \GkII}, the form of the zero-mode equations is reduced to Eq.~(\ref{eq:holomorphic_eq}) or (\ref{eq:anti-holomorphic_eq}) and the mode functions of $\psi^{(0)}_{L+}$ and $\psi^{(0)}_{L-}$ take the general holomorphic and anti-holomorphic forms, respectively.

\renewcommand{\arraystretch}{1.3}
\begin{table}[tbh]
\centering
\begin{tabular}{c|c|c} \hline
& $T^2/(Z_2 \times Z'_2)$ orbifold & Rectangle \\ \hline \hline
Boundary condition & $Z_2$ \& $Z'_2$ parities & Variational principle \\ \hline
Bulk mass & Type I, $\text{\GkII}$\,: forbidden & Type I, $\text{\GkII}$, $\text{\GkIII}$\,: allowed \\
          & Type $\text{\GkIII}$\,: allowed    & \\ \hline
Non-zero KK modes & \multicolumn{2}{|c}{$\left[ M^2 + \left( \frac{n_1 \pi}{L_1} \right)^2 + \left( \frac{n_2 \pi}{L_2} \right)^2 \right]^{1/2}$ \ $(n_{1}, n_{2} = 1,2,\cdots)$} \\ \hline
\# of zero modes & two Weyls or one Dirac & undetermined in general \\ \hline
Zero mode profile & constant & {localized with parameters} \\ \hline
\end{tabular}
\caption{{Comparison of the $T^2/(Z_2 \times Z'_2)$ orbifold and the rectangle.}}
\label{tab:rectangle_VS_Z2Z2}
\end{table}
\renewcommand{\arraystretch}{1.0}


{
Finally, we summarize and compare the situations on the rectangle and on the $T^2/(Z_2 \times Z'_2)$ in table~\ref{tab:rectangle_VS_Z2Z2}.
The properties of the KK modes, namely the spectrum and the form of the mode functions, coincide completely, while we found differences in presence of the bulk mass and zero mode properties.
Here we emphasize that the profile in Eqs.~(\ref{eq:KK_zero_L+}) and (\ref{eq:KK_zero_L}) is a specific solution in Case $\text{\GkII}$, where {more general solutions} may be possible.
Apparently on a rectangle, a wider class of solutions is realizable compared with on the $T^2/(Z_2 \times Z'_2)$ orbifold.

When we focus on Type-$\text{\GkIII}$ BCs along $y_1$ and $y_2$ directions, another important dissimilarity may happen.
On a rectangle, two sets of $S^2$ parameters $(\phi^{(')}, \theta^{(')})$, which describe $U(2)$ rotations among two 4d chiral components of a 6d Dirac fermion, are able to be introduced consistently.
Here, a 6d Dirac fermion is decomposed based on the eigenvalue $\vec{n}^{(')} \cdot \vec{\Sigma}^{(')}$ [$(\vec{n}^{(')} \cdot \vec{\Sigma}^{(')})^2 = I_8$] as
$\Psi = \mathcal{P}_{\vec{n}^{(')} \cdot \vec{\Sigma}^{(')} = +1} \Psi + \mathcal{P}_{\vec{n}^{(')} \cdot \vec{\Sigma}^{(')} = -1} \Psi$.
On the $T^2/(Z_2 \times Z'_2)$ orbifold, it would be very difficult to introduce such degrees of freedom since the following relations,
\al{
\left( \Sigma_1, \Sigma_2, \Sigma_3 \right) \Sigma^\mu &= \Sigma^\mu \left( -\Sigma_1, \Sigma_2, \Sigma_3 \right), \\
\left( \Sigma_1, \Sigma_2, \Sigma_3 \right) \Sigma^{y_1} &= \Sigma^{y_1} \left( \Sigma_1, -\Sigma_2, -\Sigma_3 \right), \\
\left( \Sigma_1, \Sigma_2, \Sigma_3 \right) \Sigma^{y_2} &= \Sigma^{y_2} \left( -\Sigma_1, \Sigma_2, -\Sigma_3 \right),
}
implies that to construct two different $Z_2$ conditions consistently may face a problem except for the three trivial cases,
\al{
{\text{i)}}   \ \vec{n}^{(')} = (1,0,0),\quad
{\text{ii)}}  \ \vec{n}^{(')} = (0,1,0),\quad
{\text{iii)}} \ \vec{n}^{(')} = (0,0,1).
}
The cases of i) and ii) {might} be inconsistent like the one discussed in section~\ref{sec:foiled_case} since two projections along $y_1$ and $y_2$ becomes different.
The case iii) is just the one that we focused on in section~\ref{sec:Z2Z2_caseIII}. 
In conclusion, a rectangle allows wider possibilities also in the choice of BCs of a 6d Dirac fermion.
}

\section{Conclusions and Discussions
\label{sec:summary}}

In this manuscript, we classified possible BCs of a 6d Dirac fermion $\Psi$ on a rectangle under the requirement that the 4d Lorentz structure is maintained, and derived the profiles of the zero modes and nonzero KK modes under the two specific boundary conditions, (i) $\Psi_{R}(x,y) = 0$ at the boundaries and (ii) $\Psi_{+}(x,y) = 0$ at the boundaries.

Here, the two BCs are a limited part of the possible configurations, where along either of the direction $y_1$ or $y_2$, three types of BCs [in Eqs~(\ref{eq:type-I_condition_6d_y1})--(\ref{eq:type-III_condition_6d_y1}) for $y_1$, in Eqs~(\ref{eq:type-I_condition_6d_y2})--(\ref{eq:type-III_condition_6d_y2}) for $y_2$] were derived, where {Type-\GkI} and {Type-\GkII} conditions discriminate 4d chirality ($R$ or $L$), while {Type-\GkIII} conditions put conditions on linear combinations of the four eigenstates of the 4d and 2d chiralities ($R\pm$, $L\pm$) at the corresponding boundaries.
{Type-\GkIII} conditions are parametrized by a position of a unit 2d sphere $S^2$, where the two specific cases $(\phi, \theta) = (0, \pi)$ and $(0,0)$ correspond to the projection condition for the {internal chirality} of $+$ and $-$, respectively.
Apparently, {Type-\GkIII} conditions do not discriminate the 4d chirality and the corresponding zero modes becomes vector-like.
Therefore, such possibilities are not suitable for regenerating the chiral structure of the SM at the zero-mode sector.
On the other hand, the emergence of such rotational parameters in BCs of a single 6d field looks nontrivial, and an exhaustive analysis of mode functions when $\phi$ and $\theta$ take generic values is of interest in a theoretical point of view.

Zero modes are distinctive in general since additional conditions are imposed [Eqs.~(\ref{eq:zero_mode_condition_Lpm}) and (\ref{eq:zero_mode_condition_Rpm}) for (i), Eqs.~(\ref{eq:zero_mode_condition_+}) and (\ref{eq:zero_mode_condition_-}) for (ii)] to the 6d Dirac equations.
In the two cases of (i) and (ii), either of the ``chiral" mode [$R$ for (i), $+$ for (ii)] is constrained by the BCs and the remaining counterparts are free from conditions on the boundaries.
Hence, corresponding zero modes are not restricted from the information on the boundaries and for them, the rectangle looks the 2d Euclidean space which is symmetric under the rotation with an axis.
We explicitly looked at the symmetry in the specific solution via the Ansatz with exponential function in Eq.~(\ref{eq:Ansatz_for_leftmode}), while this symmetry can be addressed in a generic manner by focusing on the covariance of the Dirac equation for the zero modes as discussed in appendix~\ref{appendix:spinor_rotation}.
Such emergence of rotational symmetries never occurs in 5d since one additional spacial DOF is not enough for defining rotations, which is specific in (more than or equal to) six dimensions.
Another characteristic feature of the zero modes in (i) is the number of chiral modes is two and these two modes are localized towards different directions of a rectangle.
If the Higgs vacuum expectation value are dependent on $y_1$ and $y_2$, a natural mass hierarchy is expected among the two states, where we might behold the occurrence of the two-fold degenerated states.
This direction would be interesting for addressing an origin of the number of matter generations, fermion mass hierarchies and mixing patterns simultaneously, even though three generations are impossible in 6d.
More detailed discussions would be fruitful on general aspects of such solutions, including situations more than six dimensions.

{
As we discussed in section~\ref{sec:orbifold}, we can construct the corresponding cases of Case I, $\text{\GkII}$, $\text{\GkIII}$ on a rectangle, in the language of orbifolding on {$T^2/(Z_2 \times Z'_2)$}, where the two fundamental domains [of a rectangle and $T^2/(Z_2 \times Z'_2)$] are equivalent.
A part of properties, i.e. on the nonzero KK modes, is the same, while we found differences on the zero modes and on possible BCs.
Then, we concluded that wider possibilities are realized on a rectangle, compared with on $T^2/(Z_2 \times Z'_2)$.
}

In (ii), situations are more drastic, where not only the BCs, but also the mass parameter $M$ is decoupled from the equation for determining the profile of $\Psi^{(0)}_{-}$, where the zero modes can take the generic anti-holomorphic forms, where no other properties, e.g. the number of the zero modes, are determined.
In a theoretical sense, it looks interesting since this zero modes are massive with the physical mass eigenstate $M$, while they hold such a ``conformal" property.
For this concrete case, further studies are meaningful.

Finally, let us mention that the classification of possible BCs and properties of the spectrum of a 6d Dirac fermion under the BCs can be addressed in a quantum mechanical supersymmetry (see e.g. Refs.~\cite{Cooper:1994eh,Cheon:2000tq,Lim:2005rc,Lim:2007fy} {and references in~\cite{Fujimoto:2016rbr} therein}) point of view, whose details are declared in a separate publication~\cite{Fujimoto:2016rbr}.


\section*{Acknowledgements}

We thank Tomoaki Nagasawa for discussions in the early stage of this work.
This work is supported in part by Grants-in-Aid for Scientific Research [No.~15K05055 and No.~25400260 (M.S.)] from the Ministry of Education, Culture, Sports, Science and Technology (MEXT) in Japan.
{We appreciate the anonymous Referee for giving us valuable comments.}


\appendix
\section*{Appendix}

\section{{Invariance of $S_{T^2}$ under $Z_2 \times Z'_2$}
\label{appendix:Z2Z2}}

In this appendix, we show the invariance of the action $S_{T^2}$ shown in Eq.~(\ref{eq:6d_action_T2}) under the $Z_2 \times Z'_2$ transformations which corresponds to Case $\text{\GkII}$ (discussed in section~\ref{sec:Z2Z2_caseII}) and Case $\text{\GkIII}$ (discussed in section~\ref{sec:Z2Z2_caseIII}) on a rectangle.

At first, we argue the former case, where the $Z_2 \times Z'_2$ transformation is defined in Eqs.~(\ref{eq:Z2Z2_caseII_condition_1}) and (\ref{eq:Z2Z2_caseII_condition_2}).
We focus on the three patterns of transformation of $y_1 \to -y_1$,
\al{
  \int_{-L_1}^{L_{1}} \!\!\!\! dy_{1} \int_{-L_2}^{L_{2}} \!\!\!\! dy_{2}& \,
	\overline{\Psi}_{R+}(x,y_1,y_2) \, i\Gamma^{y_1} \del_{z} \Psi_{L-}(x,y_1,y_2) \notag \\
\underbrace{\longrightarrow}_{y_1 \to -y_1} \
& \int_{-L_1}^{L_{1}} \!\!\!\! dy_{1} \int_{-L_2}^{L_{2}} \!\!\!\! dy_{2} \,
	\overline{\Psi}_{R+}(x,-y_1,y_2) \, i\Gamma^{y_1} (-\del_{\bar{z}}) \Psi_{L-}(x,-y_1,y_2) \notag \\
{\overset{(\ref{eq:Z2Z2_caseII_condition_1})}{=}}
& \int_{-L_1}^{L_{1}} \!\!\!\! dy_{1} \int_{-L_2}^{L_{2}} \!\!\!\! dy_{2} \,
	(-\overline{\Psi}_{R+}(x,y_1,y_2)) \, i\Gamma^{y_1} (-1) \del_{z} \Psi_{L-}(x,y_1,y_2) \notag \\
{=}
& \int_{-L_1}^{L_{1}} \!\!\!\! dy_{1} \int_{-L_2}^{L_{2}} \!\!\!\! dy_{2} \,
	\overline{\Psi}_{R+}(x,y_1,y_2) \, i\Gamma^{y_1} \del_{z} \Psi_{L-}(x,y_1,y_2), \\[10pt]
  \int_{-L_1}^{L_{1}} \!\!\!\! dy_{1} \int_{-L_2}^{L_{2}} \!\!\!\! dy_{2}& \,
	\overline{\Psi}_{L+}(x,y_1,y_2) \, i\Gamma^{y_1} \del_{z} \Psi_{R-}(x,y_1,y_2) \notag \\
\sim
& \int_{-L_1}^{L_{1}} \!\!\!\! dy_{1} \int_{-L_2}^{L_{2}} \!\!\!\! dy_{2} \,
	(-1) \, \overline{\del_{\bar{z}} \Psi}_{L+}(x,y_1,y_2) \, i\Gamma^{y_1} \Psi_{R-}(x,y_1,y_2) \notag \\
\underbrace{\longrightarrow}_{y_1 \to -y_1} \
& \int_{-L_1}^{L_{1}} \!\!\!\! dy_{1} \int_{-L_2}^{L_{2}} \!\!\!\! dy_{2} \,
	\overline{\del_{z} \Psi}_{L+}(x,-y_1,y_2) \, i\Gamma^{y_1} \Psi_{R-}(x,-y_1,y_2) \notag \\
{\overset{(\ref{eq:Z2Z2_caseII_condition_1})}{=}}
& \int_{-L_1}^{L_{1}} \!\!\!\! dy_{1} \int_{-L_2}^{L_{2}} \!\!\!\! dy_{2} \,
	\overline{\del_{\bar{z}} \Psi}_{L+}(x,y_1,y_2) \, i\Gamma^{y_1} \Psi_{R-}(x,y_1,y_2) \notag \\
\sim
& \int_{-L_1}^{L_{1}} \!\!\!\! dy_{1} \int_{-L_2}^{L_{2}} \!\!\!\! dy_{2} \,
	\overline{\Psi}_{L+}(x,y_1,y_2) \, i\Gamma^{y_1} \del_{z} \Psi_{R-}(x,y_1,y_2), \\[10pt]
  \int_{-L_1}^{L_{1}} \!\!\!\! dy_{1} \int_{-L_2}^{L_{2}} \!\!\!\! dy_{2}& \,
	\overline{\Psi}_{L+}(x,y_1,y_2) \, i\Gamma^{\mu} \del_{\mu} \Psi_{L+}(x,y_1,y_2) \notag \\
=
& \int_{-L_1}^{L_{1}} \!\!\!\! dy_{1} \int_{-L_2}^{L_{2}} \!\!\!\! dy_{2} \,
	\overline{\Psi}_{L+}(x,y_1,y_2) \, i\Gamma^{\mu} \del_{\mu}
	\del_{z} \frac{1}{\del_z \del_{\bar{z}}} \del_{\bar{z}}  \Psi_{L+}(x,y_1,y_2) \notag \\
\sim
& \int_{-L_1}^{L_{1}} \!\!\!\! dy_{1} \int_{-L_2}^{L_{2}} \!\!\!\! dy_{2} \,
	(-1) \overline{\del_{\bar{z}} \Psi}_{L+}(x,y_1,y_2) \, i\Gamma^{\mu} \del_{\mu}
	\frac{1}{\del_z \del_{\bar{z}}} \del_{\bar{z}}  \Psi_{L+}(x,y_1,y_2) \notag \\
\underbrace{\longrightarrow}_{y_1 \to -y_1} \
& \int_{-L_1}^{L_{1}} \!\!\!\! dy_{1} \int_{-L_2}^{L_{2}} \!\!\!\! dy_{2} \,
	(-1) \overline{\del_{z} \Psi}_{L+}(x,{-y_1},y_2) \, i\Gamma^{\mu} \del_{\mu}
	\frac{1}{\del_{\bar{z}} \del_{z}} \del_{z} \Psi_{L+}(x,{-y_1},y_2) \notag \\
{\overset{(\ref{eq:Z2Z2_caseII_condition_1})}{\sim}}
& \int_{-L_1}^{L_{1}} \!\!\!\! dy_{1} \int_{-L_2}^{L_{2}} \!\!\!\! dy_{2} \,
	\overline{\Psi}_{L+}(x,y_1,y_2) \, i\Gamma^{\mu} \del_{\mu}
	{\del_{z}} \frac{1}{\del_{\bar{z}} \del_{z}} {\del_{\bar{z}}} \Psi_{L+}(x,y_1,y_2) \notag \\
=
& \int_{-L_1}^{L_{1}} \!\!\!\! dy_{1} \int_{-L_2}^{L_{2}} \!\!\!\! dy_{2} \,
	\overline{\Psi}_{L+}(x,y_1,y_2) \, i\Gamma^{\mu} \del_{\mu} \Psi_{L+}(x,y_1,y_2),
}
where two terms connected by $\sim$ are equivalent up to surface terms by integral by parts.
{In the above, we formally inserted the identity $(\del_{z} \del_{\bar{z}})/(\del_{z} \del_{\bar{z}})$, where the commutative relation $[\del_{z}, \del_{\bar{z}}] = 0$ holds.}
We {skipped} to show the following deformation
\al{
\int_{-L_1}^{L_{1}} \!\!\!\! dy_{1} \
\underbrace{\longrightarrow}_{y_1 \to -y_1} \
\int_{L_1}^{-L_{1}} \!\!\!\!\!\! d(-y_{1})
=
\int_{-L_1}^{L_{1}} \!\!\!\! dy_{1}.
}
The terms of $\overline{\Psi}_{R-} \, i\Gamma^{y_1} \del_{\bar{z}} \Psi_{L+}$, $\overline{\Psi}_{L-} \, i\Gamma^{y_1} \del_{\bar{z}} \Psi_{R+}$, $\overline{\Psi}_{L-} \, i\Gamma^{\mu} \del_{\mu} \Psi_{L-}$ can be proved straightforwardly, while the cases of $\overline{\Psi}_{R\pm} \, i\Gamma^{\mu} \del_{\mu} \Psi_{R\pm}$ is obvious to be shown.

On the other hand in this situation linking to Case $\text{\GkII}$, the bulk mass term is not invariant as
\al{
  \int_{-L_1}^{L_{1}} \!\!\!\! dy_{1} \int_{-L_2}^{L_{2}} \!\!\!\! dy_{2}& \,
	(-M) \overline{\Psi}_{R+}(x,y_1,y_2) \Psi_{L+}(x,y_1,y_2) \notag \\
=
& \int_{-L_1}^{L_{1}} \!\!\!\! dy_{1} \int_{-L_2}^{L_{2}} \!\!\!\! dy_{2} \,
	(-M) \overline{\Psi}_{R+}(x,y_1,y_2)
	{\del_{z} \frac{1}{\del_{z} \del_{\bar{z}}} \del_{\bar{z}}}
	\Psi_{L+}(x,y_1,y_2) \notag \\
\underbrace{\longrightarrow}_{y_1 \to -y_1} \
& \int_{-L_1}^{L_{1}} \!\!\!\! dy_{1} \int_{-L_2}^{L_{2}} \!\!\!\! dy_{2} \,
	(-M) \overline{\Psi}_{R+}(x,{-y_1},y_2)
	{\del_{\bar{z}} \frac{1}{\del_{\bar{z}} \del_{z}} \del_z}
	\Psi_{L+}(x,{-y_1},y_2) \notag \\
{\overset{(\ref{eq:Z2Z2_caseII_condition_1})}{=}}
& {\int_{-L_1}^{L_{1}} \!\!\!\! dy_{1} \int_{-L_2}^{L_{2}} \!\!\!\! dy_{2} \,
	(+M) \overline{\Psi}_{R+}(x,y_1,y_2)
	\del_{\bar{z}} \frac{1}{\del_{\bar{z}} \del_{z}} \del_{\bar{z}}
	\Psi_{L+}(x,y_1,y_2)} \notag \\
=
& \int_{-L_1}^{L_{1}} \!\!\!\! dy_{1} \int_{-L_2}^{L_{2}} \!\!\!\! dy_{2} \,
	{(+M)} \overline{\Psi}_{R+}(x,y_1,y_2)
	{\frac{1}{\del_{z}} \del_{\bar{z}}}
	\Psi_{L+}(x,y_1,y_2).
}
Thus, $M$ should be zero for keeping consistency.
The discussion for the transformation of $y_2 \to -y_2$ is completely parallel to the present one and then we skip to describe.

Next, we move to the situation corresponding to Case $\text{\GkIII}$.
Almost all the calculations are simple repetitions of the above.
A notable difference is in the bulk mass term.
For $\overline{\Psi}_{-} \Psi_{-}$, the following deformation declares the invariance,
\al{
  \int_{-L_1}^{L_{1}} \!\!\!\! dy_{1} \int_{-L_2}^{L_{2}} \!\!\!\! dy_{2}& \,
	(-M) \overline{\Psi}_{-}(x,y_1,y_2) \Psi_{-}(x,y_1,y_2) \notag \\
=
& \int_{-L_1}^{L_{1}} \!\!\!\! dy_{1} \int_{-L_2}^{L_{2}} \!\!\!\! dy_{2} \,
	(-M) \overline{\Psi}_{-}(x,y_1,y_2)
	{\del_{\bar{z}} \frac{1}{\del_{\bar{z}} \del_{z}} \del_{z}}
	\Psi_{-}(x,y_1,y_2) \notag \\
\sim
& \int_{-L_1}^{L_{1}} \!\!\!\! dy_{1} \int_{-L_2}^{L_{2}} \!\!\!\! dy_{2} \,
	(-M) (-1) \overline{{\del_{z}} \Psi}_{-}(x,y_1,y_2)
	{\frac{1}{\del_{\bar{z}} \del_{z}} \del_{z}}
	\Psi_{-}(x,y_1,y_2) \notag \\
\underbrace{\longrightarrow}_{y_1 \to -y_1} \
& \int_{-L_1}^{L_{1}} \!\!\!\! dy_{1} \int_{-L_2}^{L_{2}} \!\!\!\! dy_{2} \,
	(-M) (-1) \overline{\del_{\bar{z}} \Psi}_{-}(x,{-y_1},y_2)
	\frac{1}{\del_{z} \del_{\bar{z}}} \del_{\bar{z}}
	\Psi_{-}(x,{-y_1},y_2) \notag \\
{\overset{(\ref{eq:Z2Z2_caseIII_condition_1})}{=}}
& \int_{-L_1}^{L_{1}} \!\!\!\! dy_{1} \int_{-L_2}^{L_{2}} \!\!\!\! dy_{2} \,
	(-M) (-1) \overline{{\del_{z}} \Psi}_{-}(x,y_1,y_2)
	\frac{1}{\del_{\bar{z}} \del_{z}} {\del_{z}}
	\Psi_{-}(x,y_1,y_2) \notag \\
\sim
& \int_{-L_1}^{L_{1}} \!\!\!\! dy_{1} \int_{-L_2}^{L_{2}} \!\!\!\! dy_{2} \,
	(-M) \overline{\Psi}_{-}(x,y_1,y_2)
	{\del_{\bar{z}} \frac{1}{\del_{\bar{z}} \del_{z}} \del_{z}}
	\Psi_{-}(x,y_1,y_2) \notag \\
=
& \int_{-L_1}^{L_{1}} \!\!\!\! dy_{1} \int_{-L_2}^{L_{2}} \!\!\!\! dy_{2} \,
	(-M) \overline{\Psi}_{-}(x,y_1,y_2) \Psi_{-}(x,y_1,y_2).
}
We note that to show the invariance of $\overline{\Psi}_{+} \Psi_{+}$ is straightforward.
In the present situation, corresponding to Case $\text{\GkIII}$, the bulk mass term is consistent with the $Z_2 \times Z'_2$ symmetry.
The key point is that the bulk mass term is decomposed into fields and their conjugated, namely as $\overline{\Psi} \Psi = \overline{\Psi}_{+} \Psi_{+} + \overline{\Psi}_{-} \Psi_{-}$.

\section{Comments on Zero Modes in {Case \GkII}
\label{appendix:spinor_rotation}}

In this appendix, we give comments on the zero modes in {Case \GkII}.
We remind that zero modes are described by the Dirac equations in Eqs.~(\ref{eq:6d_Weyl_eq__R}) and (\ref{eq:6d_Weyl_eq__L}) under the zero mode conditions in Eqs.~(\ref{eq:zero_mode_condition_Lpm}) and (\ref{eq:zero_mode_condition_Rpm}).
Under the present BCs~(\ref{eq:BC_case-II}), only left-handed modes can exist.
But, the following discussions are applicable for the case, ${\Psi_{L}(x,y)} = 0$ at the boundaries, where right-handed modes are allowed as zero modes.

We can show that the Dirac equations~in Eq.~(\ref{eq:6d_Weyl_eq__L}) (for left-handed zero modes) are covariant under the rotation on the $y_{1}y_{2}$-plane.
Here, we define the rotation as
\al{
\begin{pmatrix} y'_{1} \\ y'_{2} \end{pmatrix}
&=
\begin{pmatrix}
\cos{\theta} & \sin{\theta} \\
-\sin{\theta} & \cos{\theta}
\end{pmatrix}
\begin{pmatrix} y_{1} \\ y_{2} \end{pmatrix}\quad
(\theta \in \mathbb{R}),
}
where $y'_{i}\,(i=1,2)$ are rotated coordinates {and we use the generator $L_y$ defined in Eq.~(\ref{eq:6d_spinor_rotation_generator})} for the corresponding spinor rotation.
The following relations are established,
\al{
&
\left\{
\begin{array}{l}
\left( \del_{y_1} - i \del_{y_2} \right) i \Gamma^{y_1} \Psi_{L-}^{(0)}(x,y) - M \Psi_{L+}^{(0)}(x,y) = 0, \\[4pt]
\left( \del_{y_1} + i \del_{y_2} \right) i \Gamma^{y_1} \Psi_{L+}^{(0)}(x,y) - M \Psi_{L-}^{(0)}(x,y) = 0,
\end{array}
\right. 
	\label{eq:first_step} \\
\Rightarrow &
\left\{
\begin{array}{l}
\left( \del_{y'_1} - i \del_{y'_2} \right) i \Gamma^{y_1} \Psi_{L-}^{(0)}(x,y') - M \Psi_{L+}^{(0)}(x,y') = 0, \\[4pt]
\left( \del_{y'_1} + i \del_{y'_2} \right) i \Gamma^{y_1} \Psi_{L+}^{(0)}(x,y') - M \Psi_{L-}^{(0)}(x,y') = 0,
\end{array}
\right. \notag \\
\Leftrightarrow &
\left\{
\begin{array}{l}
\left( \del_{y_1} - i \del_{y_2} \right) i \Gamma^{y_1} \left( e^{i\theta} \Psi_{L-}^{(0)}(x,y') \right) - M \Psi_{L+}^{(0)}(x,y') = 0, \\[4pt]
\left( \del_{y_1} + i \del_{y_2} \right) i \Gamma^{y_1} \Psi_{L+}^{(0)}(x,y') - M \left( e^{i\theta} \Psi_{L-}^{(0)}(x,y') \right) = 0,
\end{array}
\right.
}
where the first manipulation is a simple reparametrization about $y_1$ and $y_2$, and we use the property $e^{-i\theta L_y} \Gamma^{y_1} = \Gamma^{y_1} e^{+i\theta L_y}$ which is easily shown from Eq.~(\ref{eq:Gamma_Clifford}).
The above sequence implies that if $\{ \Psi^{(0)}_{L+}(x,y), \Psi^{(0)}_{L-}(x,y) \}$ is a set of solutions of the equations in Eq.~(\ref{eq:first_step}) [or (\ref{eq:6d_Weyl_eq__L})], the set $\{ \Psi^{(0)}_{L+}(x,y'), e^{i\theta} \Psi^{(0)}_{L-}(x,y') \}$ also acts as a set of solutions.
We can check this relationship by use of a concrete example.
When we start with a special solution of one in Eq.~(\ref{eq:KK_zero_L}) by setting $j=1$ and $\theta = 0$,
\al{
\Psi_{L+}^{(0)}(x,y) &=  N \, \xi^{(0)}_{L}(x) \, e^{M y_1}, \\
\Psi_{L-}^{(0)}(x,y) &=  N (i \Gamma^{y_1}) \, \xi^{(0)}_{L}(x) \, e^{M y_1},
}
the corresponding set of solutions takes the forms,
\al{
\Psi_{L+}^{(0)}(x,y) &= N \, \xi^{(0)}_{L}(x) \, e^{M(\cos{\theta} y_1 + \sin{\theta} y_2)}, \\
\Psi_{L-}^{(0)}(x,y) &= N \, e^{i\theta} (i \Gamma^{y_1}) \, \xi^{(0)}_{L}(x) \, e^{M(\cos{\theta} y_1 + \sin{\theta} y_2)},
} 
which are nothing but a general solution in Eq.~(\ref{eq:KK_zero_L}) when the angle $\theta$ is real.
This property manifestly shows the set of solutions being symmetric under the rotation.

We note that in the solution~(\ref{eq:KK_zero_L}), the parameter $\theta$ can be complex, although $\theta$ should be real for rotations.
This is because the derivation in Eq.~(\ref{eq:KK_zero_L}) holds even for complex $\theta$.
This may be called a complexification of the rotation.

\section{{Physics of the angle $\theta$ as mass hierarchy} \label{sec:theta}}

In this {appendix}, we show that the parameter $\theta$ in zero modes~(\ref{eq:KK_zero_L_1}), (\ref{eq:KK_zero_L_2}) is not {an unphysical} parameter but {physical one, which affects the actual physical values, e.g., 4d masses of zero modes} through Yukawa couplings.
Imitating the structure of {Yukawa couplings in} the {standard model~(SM)},
we demonstrate the physical implication of the parameter $\theta$ {in a toy example}.
To this end, let us consider the following Yukawa interaction term with two 6d Dirac fermions $\Psi'(x,y)$, $\Psi(x,y)$ and a VEV $\langle H(x,y)\rangle$ of a 6d scalar field:
	\al{
	S^{({\rm Y})}=\int \! d^{4} x \int_{0}^{L_{1}} \!\!\!\! dy_{1} \int_{0}^{L_{2}} \!\!\!\! dy_{2} \,
	&\left[\overline{\Psi}'(x,y) \left( i \Gamma^A \del_A - M' \right) \Psi'(x,y)+\overline{\Psi}(x,y) \left( i \Gamma^A \del_A - M \right) \Psi(x,y)\right.\nonumber\\
	&\left.\hspace{2em}+\lambda\, \overline{\Psi}' (x,y)\langle H^{{\ast}}(x,y) \rangle\Psi(x,y)+(\text{h.c.})\right].\label{S-Yukawa}
	}
{We find resemblance in $\Psi'$, $\Psi$ and $H$ to} a 4d right-handed chiral fermion, a 4d left-handed chiral fermion and the Higgs field $H(x)$ in the {SM}.
{The complex conjugation in the VEV reflects the correspondence in gauge structure to the SM.}
Since $\Psi'$ ($\Psi$) is an imitation of a 4d right-handed (left-handed) chiral fermion, we impose the following BCs.
	\al{
	\mathcal{P}_L \Psi'(x,y) = 0\qquad \text{at} \quad y_{1}=0,L_{1},\ y_2 = 0,L_2.\\
	\mathcal{P}_R \Psi(x,y) = 0\qquad \text{at} \quad y_{1}=0,L_{1},\ y_2 = 0,L_2.	
	}
Thanks to the above BCs, we obtain the twofold degenerated chiral zero modes $\xi'{}^{(0)}_{R1}(x),\xi'{}^{(0)}_{R2}(x)$ for $\Psi'$ and  $\xi^{(0)}_{L1}(x),\xi^{(0)}_{L2}(x)$ for $\Psi$ as a {toy example} of the {SM}.
As expressed in {Eqs.}~(\ref{eq:KK_zero_L_1})-(\ref{eq:KK_zero_L_2}), the explicit form of the chiral zero modes are given {as follows:}
	\al{
	&\Psi'(x,y)\supset \Psi'{}_{R+}^{(0)}(x,y)+\Psi'{}_{R-}^{(0)}(x,y),\label{eq:Psi'-zeros}\\
	&\Psi'{}_{R+}^{(0)}(x,y)=
	N'_{1}\, \xi'{}_{R1}^{(0)}(x)e^{M'(\cos\theta'\,y_{1}+\sin\theta'\,y_{2})}
	+N'_{2}\, \xi'{}_{R2}^{(0)}(x)e^{-M'(\cos\theta'{}^{\ast}\,y_{1}+\sin\theta'{}^{\ast}\,y_{2})},\label{eq:KK-zero-R1}\\
	&\Psi'{}_{R-}^{(0)}(x,y)=
	N'_{1}\, e^{i\theta'}(i\Gamma^{y_1}) \, \xi'{}_{R1}^{(0)}(x)e^{M'(\cos\theta'\,y_{1}+\sin\theta'\,y_{2})}
	-N'_{2}\, e^{i\theta'{}^{\ast}}(i\Gamma^{y_1}) \, \xi'{}_{R2}^{(0)}(x)e^{-M'(\cos\theta'{}^{\ast}\,y_{1}+\sin\theta'{}^{\ast}\,y_{2})},\label{eq:KK-zero-R2}\\
	&\Psi(x,y)\supset \Psi_{L+}^{(0)}(x,y)+\Psi_{L-}^{(0)}(x,y),\\
	&\Psi_{L+}^{(0)}(x,y) = 
	N_1 \, \xi^{(0)}_{L1}(x) \, e^{M(\cos{\theta} y_1 + \sin{\theta} y_2)} +
	N_2 \, \xi^{(0)}_{L2}(x) \, e^{-M(\cos{\theta^\ast} y_1 + \sin{\theta^\ast} y_2)}, \label{eq:KK-zero-L1}\\
	&\Psi_{L-}^{(0)}(x,y) = 
	N_1 \, e^{i\theta} (i\Gamma^{y_1}) \, \xi^{(0)}_{L1}(x) \, e^{M(\cos{\theta} y_1 + \sin{\theta} y_2)} -
	N_2 \, e^{i\theta^\ast} (i\Gamma^{y_1}) \, \xi^{(0)}_{L2}(x) \, e^{-M(\cos{\theta^\ast} y_1 + \sin{\theta^\ast} y_2)},
	\label{eq:KK-zero-L2}
	}
where $N'_{1}$, $N'_{2}$, $N_{1}$, $N_{2}$ are normalization factors given by 
	\al{
	N'_{1}&=\sqrt{\frac{M'{}^{2}(\cos\theta'{}^{\ast}+\cos\theta')(\sin\theta'{}^{\ast}+\sin\theta')}{(1+e^{{-}i(\theta'{}^{\ast}-{\theta'})})(e^{{M'}(\cos\theta'{}^{\ast}+{\cos\theta'})L_{1}}-1)(e^{{M'}(\sin\theta'{}^{\ast}+\sin\theta')L_{2}}-1)}},\\
	N'_{2}&=\sqrt{\frac{M'{}^{2}(\cos\theta'{}^{\ast}+\cos\theta')(\sin\theta'{}^{\ast}+\sin\theta')}{(1+e^{{+}i(\theta'{}^{\ast}-{\theta'})})(1-e^{-{M'}(\cos\theta'{}^{\ast}+{\cos\theta'})L_{1}})(1-e^{-{M'}(\sin\theta'{}^{\ast}+\sin\theta')L_{2}})}},\\
	N_1 &=
	\sqrt{\frac{M^2 (\cos{\theta^\ast} + \cos{\theta}) (\sin{\theta^\ast} + \sin{\theta})}
	{(1 + e^{-i(\theta^\ast - \theta)})(e^{M(\cos{\theta^\ast} + \cos{\theta})L_1} -1)
	 (e^{M(\sin{\theta^\ast} + \sin{\theta})L_2} -1)}}, \\
	N_2 &=
	\sqrt{\frac{M^2 (\cos{\theta^\ast} + \cos{\theta}) (\sin{\theta^\ast} + \sin{\theta})}
	{(1 + e^{+i(\theta^\ast - \theta)})(1 - e^{-M(\cos{\theta^\ast} + \cos{\theta})L_1})
	 (1 - e^{-M(\sin{\theta^\ast} + \sin{\theta})L_2})}}.	
	}
We now assume that the VEV of the 6d scalar field $\langle H(x,y)\rangle$ has a form of 
	\al{
	\langle H(x,y)\rangle=v\,e^{M_{H}y_{1}},\label{HiggsVEV}
	}
where $M_{H}$ is a parameter which possesses mass-dimension {one} and the constant $v$ possesses a mass-dimension {two}. 
We shall give some comments for the above VEV of the scalar field. In the context of higher-dimensional theory, it is known that the extra-dimension coordinate-dependent VEV of the {scalar field} gives a chance to solve the fermion mass hierarchy problem through {an} overlap integral with respect to the extra dimension \cite{Kaplan:2001ga}. Moreover, it was also unveiled that a VEV of a scalar field inevitably possesses an extra-dimension coordinate-dependence when we consider a general class of BCs \cite{Fujimoto:2011kf}. Thus in a general framework, as we consider in this paper, the form of the VEV (\ref{HiggsVEV}) is expected to be realized easily and convenient to discuss physics, e.g., 4d masses of zero modes and a mass hierarchy of them, though the form of the VEV~(\ref{HiggsVEV}) is not essential and other $y$-dependent forms may work well, too.

From the forms (\ref{eq:KK-zero-R1}){,} (\ref{eq:KK-zero-R2}), (\ref{eq:KK-zero-L1}){,} (\ref{eq:KK-zero-L2}) and (\ref{HiggsVEV}), we can read that $\xi'{}_{R1}^{(0)}$ and $\xi_{L1}^{(0)}$ localize to the direction
{$y' \equiv M'(\cos{\theta'} \, y_{1}+\sin\theta' \, y_{2})$}
and
{$y \equiv M(\cos{\theta} \, y_{1}+\sin\theta \, y_{2})$.}
On the other hand, $\xi'{}_{R2}^{(0)}$ and $\xi_{L2}^{(0)}$ localize to the direction $-y'$ and $-y$.
{The} fact that twofold degenerated zero modes possess different localization {directions} with each other plays an important role when we discuss 4d masses of them. We put a localization direction of the 6d scalar VEV $\langle H(x,y)\rangle$ as $y_{1}$-direction for simplicity.

Substituting the forms (\ref{eq:Psi'-zeros})-(\ref{eq:KK-zero-L2}) and (\ref{HiggsVEV}) into {Eq.}~(\ref{S-Yukawa}), we can derive the following action for the zero-mode part:
	\al{
	S^{({\rm Y})}|_{\text{zero-mode part}} &= \int d^4 x \
\Bigg\{\sum^{2}_{j=1}\overline{\xi}'{}^{(0)}_{Rj}(x) i \Gamma^\mu \del_\mu \xi'{}_{Rj}^{(0)}(x)+\sum^{2}_{k=1}\overline{\xi}^{(0)}_{Lk}(x) i \Gamma^\mu \del_\mu \xi_{Lk}^{(0)}(x)\nonumber\\
	&\hspace{12em} +\sum^{2}_{j=1}\sum^{2}_{k=1}m_{jk}\,\overline{\xi}'{}^{(0)}_{Rj}(x) \xi_{Lk}^{(0)}(x)+(\text{h.c.})
\Bigg\},	}
where
	\al{
	m_{11}&=\lambda v\, N'_{1}N_{1}
	(1-e^{-i(\theta'{}^{\ast}-\theta)})
	\left(\frac{e^{(M_{H}+M'\cos\theta'{}^{\ast}+M\cos\theta)L_{1}}-1}{M_{H}+M'\cos\theta'{}^{\ast}+M\cos\theta}\right)
	\left(\frac{e^{(M'\sin\theta'{}^{\ast}+M\sin\theta)L_{2}}-1}{M'\sin\theta'{}^{\ast}+M\sin\theta}\right),\\
	m_{12}&=\lambda v\, N'_{1}N_{2}
	(1+e^{-i(\theta'{}^{\ast}-\theta^{\ast})})	
	\left(\frac{e^{(M_{H}+M'\cos\theta'{}^{\ast}-M\cos\theta^{\ast})L_{1}}-1}{M_{H}+M'\cos\theta'{}^{\ast}-M\cos\theta^{\ast}}\right)
	\left(\frac{e^{(M'\sin\theta'{}^{\ast}-M\sin\theta^{\ast})L_{2}}-1}{M'\sin\theta'{}^{\ast}-M\sin\theta^{\ast}}\right),\label{m12}\\
	m_{21}&=\lambda v\, N'_{2}N_{1}
	(1+e^{-i(\theta'-\theta)})
	\left(\frac{e^{(M_{H}-M'\cos\theta'+M\cos\theta)L_{1}}-1}{M_{H}-M'\cos\theta' +M\cos\theta}\right)
	\left(\frac{e^{(-M'\sin\theta' +M\sin\theta)L_{2}}-1}{-M'\sin\theta' +M\sin\theta}\right),\label{m21}\\
	m_{22}&=\lambda v\, N'_{2}N_{2}
	(1-e^{-i(\theta'-\theta^{\ast})})
	\left(\frac{e^{(M_{H}-M'\cos\theta' -M\cos\theta^{\ast})L_{1}}-1}{M_{H}-M'\cos\theta'-M\cos\theta^{\ast}}\right)
	\left(\frac{1-e^{-(M'\sin\theta'+M\sin\theta^{\ast})L_{2}}}{M'\sin\theta' +M\sin\theta^{\ast}}\right).
	}
{Obviously} $m_{jk}$ ($j=1,2$; $k=1,2$) depends on the parameter $\theta$ and the degeneracy of zero modes {can be} resolved. Therefore we {conclude} that the parameter $\theta$, which appears in zero mode solutions, actually affects to the physical quantities.

To show the possibility to solve the fermion mass hierarchy by use of this physical parameter $\theta$, let us consider the following special choice:
	\al{
	&\theta'=0,\quad \theta+\pi=0,\quad M=-M'{,}
	\label{examine-choice}
	}
as an illustrative example.\footnote{
{It is noted that under the specific condition on the angles, $\theta' = \theta + \pi$ {($\theta, \theta' \in \mathbb{R}$)}, the two limited configurations, $M_H = 0$ or $M' = M$, lead to degenerated mass spectra, {and also that a hierarchical spectrum can be obtained for $\theta' = \theta + \pi/2$, independently of the value of $\theta$ with proper choices of the parameters.}
The authors thank the Referee for pointing out {these properties}.}
}
This parameter choice {shows} the possibility to solve the fermion mass hierarchy. Under the choice of the parameters, zero modes of the 6d Dirac fermions $\Psi'$ and $\Psi$ are expressed as 
	\al{
	\Psi'(x,y)&\supset \Psi'{}^{(0)}_{R+}(x,y)+\Psi'{}^{(0)}_{R-}(x,y)\nonumber\\
	&={\sqrt{\frac{M}{(1-e^{-2ML_{1}})L_2}}}\Bigl(\xi'{}^{(0)}_{R1}(x)+ i\Gamma^{y_{1}}\xi'{}^{(0)}_{R1}(x)\Bigr)e^{-M y_{1}} + {\sqrt{\frac{M}{(e^{2ML_{1}}-1)L_2}}} \Bigl(\xi'{}^{(0)}_{R2}(x)- i\Gamma^{y_{1}}\xi'{}^{(0)}_{R2}(x)\Bigr)e^{M y_{1}},\\
	\Psi(x,y)&\supset \Psi^{(0)}_{L+}(x,y)+\Psi'^{(0)}_{L-}(x,y)\nonumber\\
	&= {\sqrt{\frac{M}{(1 - e^{-2ML_{1}})L_2}}} \Bigl(\xi^{(0)}_{L1}(x)-i\Gamma^{y_{1}}\xi^{(0)}_{L1}(x)\Bigr)e^{{-M} y_{1}} + {\sqrt{\frac{M}{(e^{2ML_{1}} -1)L_2}}} \Bigl(\xi^{(0)}_{L2}(x)+ i\Gamma^{y_{1}}\xi^{(0)}_{L2}(x)\Bigr)e^{{M} y_{1}}
	}
This expansion leads us to the results,
	\al{
	S^{({\rm Y})}|_{\text{zero-mode part}} &= \int d^4 x \
\Bigg\{\sum^{2}_{j=1}\overline{\xi}'{}^{(0)}_{Rj}(x) i \Gamma^\mu \del_\mu \xi'{}_{Rj}^{(0)}(x)+\sum^{2}_{k=1}\overline{\xi}^{(0)}_{Lk}(x) i \Gamma^\mu \del_\mu \xi_{Lk}^{(0)}(x)\nonumber\\
	&\hspace{12em} +\sum^{2}_{j=1}m_{jj}\,\overline{\xi}'{}^{(0)}_{Rj}(x) \xi_{Lj}^{(0)}(x)+(\text{h.c.})
\Bigg\},	}
where
	\al{
	m_{11}&\simeq \lambda v\,  \left(\frac{2M}{M_{H}-2M}\right)e^{(M_{H}-2M) L_{1}},\\
	m_{22}&\simeq \lambda v\,  \left(\frac{2M}{M_{H}+2M}\right)e^{{M_{H}} L_{1}},
	}
{where we note that the factor $\lambda v$ has mass dimension one.}
In the above calculation, we introduced approximations $(M_{H}\pm 2M)L_{1}\gg 1$ and $ML_{1}\gg 1$ for convenience. We can easily find that a mass hierarchy $m_{22}\gg m_{11}$ appears to the two-generation fermions since the ratio of the masses are given as
	\al{
	\frac{m_{22}}{m_{11}}\simeq \left(\frac{M_{H}-2M}{M_{H}+2M}\right)e^{{2ML_{1}}}.
	}
	So we can conclude that a mass hierarchy appears in the 4d masses of zero modes with introducing an extra-dimension coordinate-dependent VEV of the scalar since the parameter $\theta$ control a localization direction of zero modes and twofold degenerated zero modes possess different localization {directions} with each other.

Finally, we give a comment for a flavor mixing. In this special parameter choice, off diagonal components of the mass matrix{,} $m_{12}$, $m_{21}$ vanish. However, as we can see in {Eqs.}~(\ref{m12}) {and} (\ref{m21}), off diagonal components appear naturally in {a} general choice of the parameters so that a flavor mixing can occur naturally in the case of general choices.

\bibliographystyle{utphys}
\bibliography{draft_6dDirac_modefunction}

\end{document}